\documentclass[twocolumn,showpacs,preprintnumbers,showkeys,superscriptaddress,floatfix]{revtex4}  
\usepackage{amssymb}
\usepackage{graphicx}
\usepackage{dcolumn}
\usepackage{bm}
\usepackage{appendix}
\usepackage{epstopdf}

\begin{document}

\title{Generalized-Seniority Pattern and Thermal Properties in Even Sn Isotopes}
\author{L. Y. Jia}  \email{liyuan.jia@usst.edu.cn}
\affiliation{Department of Physics, University of Shanghai for
Science and Technology, Shanghai 200093, P. R. China}
\author{Chong Qi}
\affiliation{Department of Physics, Royal Institute of Technology (KTH), SE-10691 Stockholm, Sweden}

\date{\today}

\begin{abstract}

Even tin isotopes of mass number $A = 108 \sim 124$ are calculated with realistic interactions in the generalized-seniority approximation of the nuclear shell model. For each nucleus, we compute the lowest ten thousand states ($5000$ of each parity) up to around $8$ MeV in excitation energy, by allowing as many as four broken pairs. The lowest fifty eigen energies of each parity are compared with the exact results of the large-scale shell-model calculation. The wavefunctions of the mid-shell nuclei show a clear pattern of the stepwise breakup of condensed coherent pairs with increasing excitation energy. We also compute in the canonical ensemble the thermal properties -- level density, entropy, and specific heat -- in relation to the thermal pairing phase transition.
\end{abstract}


\vspace{0.4in}

\maketitle


\section{Introduction}

The similarity between metal superconductivity and nuclear superfluidity was soon recognized \cite{Bohr_1958, Belyaev_1959} after the highly successful BCS theory \cite{BCS}. In the superconducting ground state electrons are coupled into pairs by the lattice-mediated effective attraction. In nuclei the short-range pairing force tends to couple nucleons. Although the pairing force influences practically all nuclei across the nuclear chart, the best superfluids are found among spherical semi-magic and near-magic nuclei. In these nuclei, the pairing force dominates over other correlations, including the most important quadrupole-quadrupole one. As a result, the ground state is largely a condensate of coherent pairs, well separated from the rest states by the pairing gap.

Naturally, we would ask to what extend this superfluid structure persists at higher energies in the increasingly dense spectrum. In a purely pairing model, the states with $s$ broken-pairs are roughly degenerate at $s$ times the pairing gap. But in reality other correlations may disturb this picture.

In the standard BCS and Bogoliubov quasi-particle formalism, one may consider diagonalizing $\hat{H} - \lambda \hat{N}$ ($\lambda$ is the chemical potential) in the subspaces with two or more quasi-particles until convergence. However, the eigen wavefunctions -- even for the lowest states -- usually have a large number of quasi-particles thus converge slowly (see Ref. \cite{Allaart_1988} and references therein). This is due to the particle-number violating nature of the quasi-particle formalism. Instead the most natural approach seems to be the generalized-seniority truncation of the shell-model. The starting point is the fully paired state [Eq. (\ref{gs}) below] that is the trial BCS wavefunction projected onto the good particle number. The coherent pair structure is usually fixed by minimizing the mean energy in the variation principle. As a truncation scheme of the shell-model, we break the coherent pairs and diagonalize the Hamiltonian $H$ in the subspace consisting of all the states with $S=2s$ unpaired nucleons. [The subspace of $S$ unpaired nucleons consists of the subspaces of generalized-seniority $S$, $S-2$, ..., $2$, $0$; see Eq. (\ref{le_s_space}).] Increasing $S$ thus the subspace size, the eigen wavefunctions gradually converge to the exact shell-model ones when all the pairs are broken. The distribution in terms of generalized seniority is given by the amplitudes of various $S$ components in the eigen wavefunction.

The long chain of tin isotopes is a classical example for nuclear superfluidity and attracts continuous attention. Recent theoretical investigations include for examples the mean-field theory \cite{Shimoyama_2011, Grasso_2012, Potel_2011, Terasaki_2005, Niksic_2002, Li_2015}, the nuclear shell model \cite{Dean_2003, Brown_2005, Iskra_2014, Qi_2012}, and the schematic generalized seniority description \cite{Morales_2011, Maheshwari_2016, Maheshwari_2016_2}. The near constancy of the first $2^+$ excitation energy suggests strong pairing correlations and leads Talmi \cite{Talmi_1971, Talmi_book, Morales_2011} to propose the generalized seniority as a good quantum number of the Hamiltonian. Realistic Hamiltonians may not satisfy the derived restrictions, and the universal method should be the diagonalization of the Hamiltonian in the subspace of $S$ unpaired nucleons (the subspace up to generalized-seniority $S$). This has been done in the neutron $50 \sim 82$ major shell with either phenomenological or more realistic interactions for $S = 2$ \cite{Ottaviani_1969, Gambhir_1971} and $S = 4$ \cite{Bonsignori_1985, Allaart_1988}, but with modern realistic interactions \cite{Jensen_1995} only for $S = 2$ \cite{Sandulescu_1997, Dean_2003}. However, as pointed out in these works, higher $S$ is necessary to achieve convergence, which is challenging and not done yet. Using the fast algorithm for generalized seniority we developed \cite{Jia_2015} and applied \cite{Qi_2016} recently, in this work we
compute with modern realistic interactions \cite{Qi_2012} the lowest ten thousand states ($5000$ of each parity) of $^{108 \sim 124}$Sn in the $S = 8$ subspace. The distribution of wavefunctions in terms of generalized seniority is examined in detail. The results show that the superfluid structure indeed persists in the higher dense spectrum, and the picture of successive breakup of coherent pairs is approximately valid, especially so near the mid-shell.

The thermal properties of tin isotopes related to the pairing phase transition were the subject of many recent studies. The $J$-level densities up to the neutron separation energy in $^{116,118,122}$Sn have been extracted experimentally under certain assumptions \cite{Agvaanluvsan_2009, Toft_2010, Toft_2011}, together with the specific heat and entropy. Theoretically, the thermal properties could be computed by the finite-temperature mean-field theory \cite{Sano_1963, Goodman_1981, Li_2015}. When practical, the spectroscopically accurate shell-model type approaches are possibly more accurate. The shell model Monte Carlo method \cite{Lang_1993, Alhassid_1994} is suitable by its formulation to compute canonical-ensemble properties \cite{Liu_2001, Alhassid_2003}, but is limited by the sign problem. The standard shell model is able to compute couples of the low-lying states but is still time-consuming near the mid-shell \cite{Qi_2012}. Accurate thermal properties in the canonical ensemble require usually thousands of states and few calculations have been done in this way \cite{Shimizu_2016}. In this work we are able to compute the lowest ten thousand states owing to the relatively small dimension of the generalized-seniority truncated subspace ($S=8$); this is enough for the accurate computation of canonical-ensemble quantities up to a temperature $k_B T \approx 0.5$ MeV. The transition from the superfluid phase to the normal phase is reproduced.

Section \ref{sec_formalism} briefly reviews the generalized-seniority formalism. The results for tin isotopes are presented in Sec. \ref{sec_Pattern} inspecting the generalized-seniority structure of the eigen wavefunctions. Based on the spectrum, we compute canonical-ensemble thermal properties in Sec. \ref{sec_thermal} in relation to the pairing phase transition. Section \ref{sec_conclusions} summarizes the work.


\section{Generalized Seniority Formalism  \label{sec_formalism}}

We briefly review the generalized-seniority formalism in relation to the current work. The
pair-creation operator
\begin{eqnarray}
P_\alpha^\dagger = a_\alpha^\dagger a_{\tilde{\alpha}}^\dagger  \label{P1_dag}
\end{eqnarray}
creates a pair of particles on the single-particle level $|\alpha\rangle$
and its time-reversed partner $|\tilde{\alpha}\rangle$
($|\tilde{\tilde{\alpha}}\rangle = - |\alpha\rangle$, $P_\alpha^\dagger = P_{\tilde{\alpha}}^\dagger$). The coherent
pair-creation operator
\begin{eqnarray}
P^\dagger = \sum_{m_\alpha > 0} v_\alpha P_\alpha^\dagger  \label{P_dag}
\end{eqnarray}
creates a pair of particles coherently distributed with structure
coefficients $v_\alpha$ over the entire single-particle space, where the summation runs over
orbits with a positive magnetic quantum
number $m_\alpha$. The pair-condensate wavefunction of the $2N$-particle system
\begin{eqnarray}
(P^\dagger)^{N} |{\rm vac}\rangle
 \label{gs}
\end{eqnarray}
builds in pairing correlations, where $|{\rm vac}\rangle$ is the vacuum state.

Gradually breaking coherent pairs, the state with $S=2s$ unpaired nucleons is
\begin{eqnarray}
\underbrace{a^\dagger a^\dagger ... a^\dagger}_{S =
2s} (P^\dagger)^{N-s} | 0 \rangle .
\label{sen_basis}
\end{eqnarray}
Loosely speaking, $S$ is defined as the generalized-seniority quantum number \cite{Talmi_1971, Shlomo_1972, Gambhir_1969, Gambhir_1971, Allaart_1988, Talmi_book}. More precisely, we distinguish between the space $|S\}$ of $S$ unpaired nucleons and the space $|S\rangle$ of generalized-seniority $S$. The space $|S\}$ consists of all the states of the form (\ref{sen_basis}). Any state of $S' < S$ unpaired nucleons can be written as a linear combination of the states of $S$ unpaired nucleons, after substituting several $P^\dagger$ by Eq. (\ref{P_dag}). Therefore $|S'\}$ is a subspace of $|S\}$,
\begin{eqnarray}
| S \} \supset |S-2\}  \supset |S-4\} \supset ... \supset |2\} \supset |0\} .  \label{S_link_bp}
\end{eqnarray}
In contrast, $|S\rangle$ is the subspace after removing the subspace $|S-2\}$ from the space $|S\}$, thus
\begin{eqnarray}
&~& | S \}  \nonumber \\
&=& |S\rangle \cup | S-2 \}  \nonumber \\
&=& |S\rangle \cup | S-2 \rangle \cup |S-4\}  \nonumber \\
&=& ... \nonumber \\
&=& |S\rangle \cup |S-2\rangle \cup ... \cup |2\rangle \cup |0\rangle .  \label{le_s_space}
\end{eqnarray}
The symbol ``$\cup$'' means set union. In this work $S=2s$ is even, and we define $|s\} \equiv |S\}$ and $|s\rangle \equiv |S\rangle$. The original basis vectors (\ref{sen_basis}) are not orthogonal. After orthonormalization the new basis vectors of the space $|s\rangle$ are enumerated as $|s,i\rangle$, where the index $i$ runs from one to the dimension of $|s\rangle$.

Practical generalized-seniority calculations usually truncate the full many-body space to the subspace $|s\}$ and then diagonalize the Hamiltonian ($s = N$ corresponds to the full space without truncation). The eigen wavefunction is
\begin{eqnarray}
| E \rangle = \sum_{s' \le s} \sum_i c_{s',i} |s',i\rangle .  \label{E_wf}
\end{eqnarray}
Investigating the wavefunction (\ref{E_wf}) in terms of generalized seniority, the amplitude for generalized-seniority $2s'$ is
\begin{eqnarray}
P(s') = \sum_i |c_{s',i}|^2 .  \label{P_s}
\end{eqnarray}
And $\sum_{s' \le s} P(s') = 1$. The average of $s$ is
\begin{eqnarray}
\bar{s} = \sum_{s' \le s} s' P(s') .  \label{s_bar}
\end{eqnarray}
The average of $s^2$ is
\begin{eqnarray}
\overline{s^2} = \sum_{s' \le s} (s')^2 P(s') .  \label{s2_bar}
\end{eqnarray}
The fluctuation of $s$ is
\begin{eqnarray}
\Delta s = \sqrt{ \overline{(s - \bar{s})^2} } = \sqrt{ \overline{s^2} - (\bar{s})^2 } .  \label{Delta_s}
\end{eqnarray}


\section{Generalized-Seniority Pattern  \label{sec_Pattern}}

In this work even tin isotopes of mass number $A = 108 \sim 124$ are computed in the generalized-seniority truncation of the shell model to four broken pairs (up to generalized-seniority eight). The doubly magic $^{100}_{~50}$Sn is taken as an inert core and the valence neutrons distribute in the $50 \sim 82$ major shell. We take the Hamiltonian from Ref. \cite{Qi_2012}. It starts from the realistic CD-Bonn nucleon-nucleon potential \cite{ref_CD_Bonn} and is renormalized in the perturbative $G$-matrix approach \cite{Jensen_1995}. Then the monopole terms and unknown single-particle energies are fitted \cite{Qi_2012} to the $157$ experimental low-lying yrast energies in $^{102 \sim 132}$Sn of both even and odd masses. This Hamiltonian has been used in Refs.
\cite{Back_2013, Procter_2013, Jiao_2014, Doncel_2015, Neacsu_2016, Horoi_2016}.

\begin{figure}
\includegraphics[width = 0.45\textwidth]{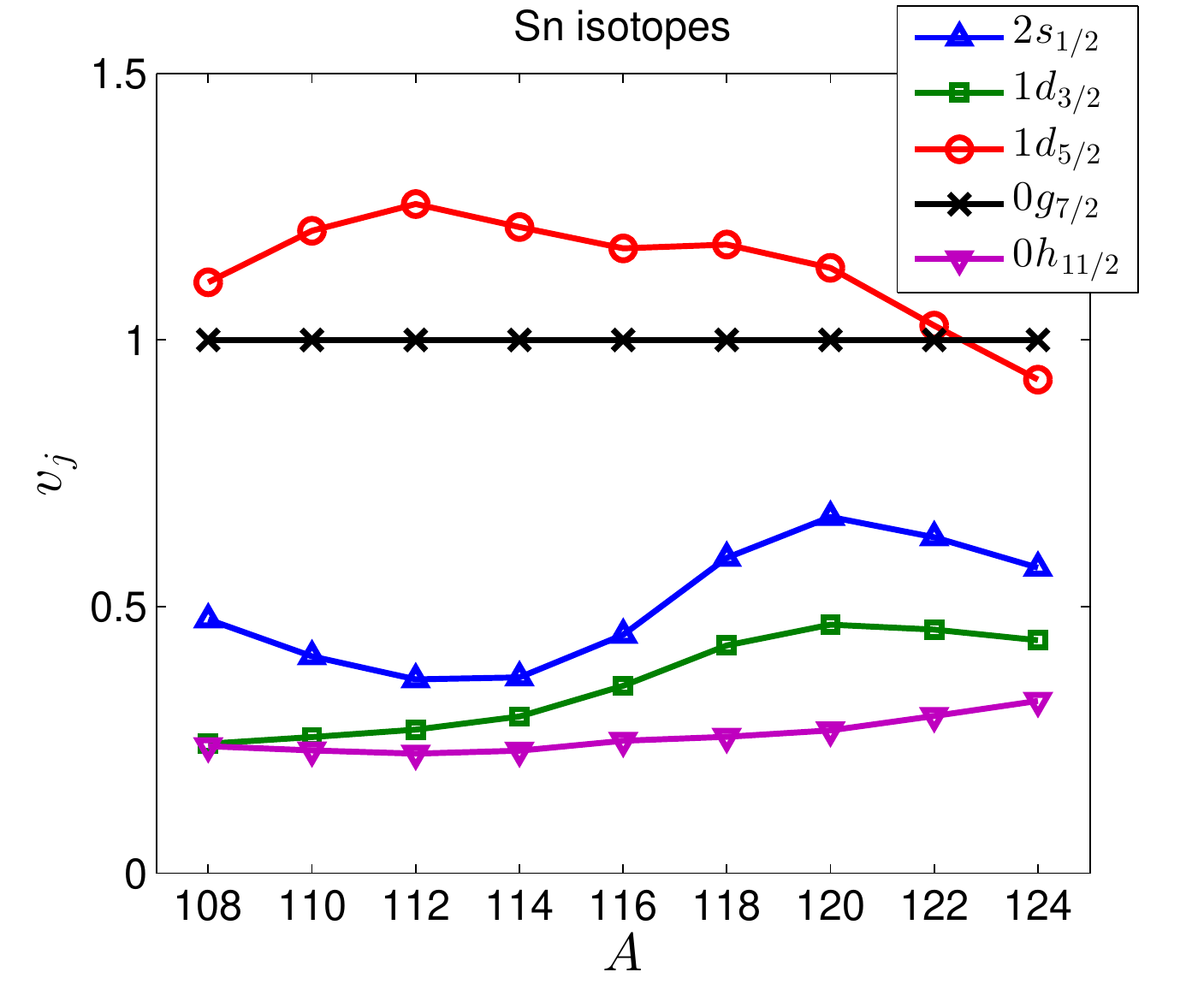}
\caption{\label{Fig_v_in} (Color online)  Collective pair structures $v_j$ (\ref{P_dag}) in tin isotopes ($A$ is the mass number). They are normalized such that $v_{0g_{7/2}} = 1$. }
\end{figure}

For each nucleus we determine the coherent pair structure $v_j$ (\ref{P_dag}) in the variation principle through minimizing the mean energy of the fully paired state [Eq. (\ref{gs}), $S=0$]. This state has the particle-hole symmetry \cite{Talmi_1982}; the energy minimum and the wavefunction (\ref{gs}) are independent of whether choosing the particles or the holes as the degree of freedom. The particle-pair structures and the hole-pair structures are reciprocals of each other [$v_j^{\rm{hole}} = 1/v_j^{\rm{particle}}$ in Eq. (\ref{P_dag})]. Figure \ref{Fig_v_in} shows the particle-pair structures $v_j^{\rm{particle}}$ in $^{108 \sim 124}$Sn normalized by setting $v_{0g_{7/2}}^{\rm{particle}} = 1$. In the isotopic chain $v_j$'s vary smoothly with the mass number. The two orbits $0g_{7/2}$ and $1d_{5/2}$ that are lower in energy have larger $v_j$ and bigger occupancy.

In this work we truncate the full shell-model space to the subspace of four broken pairs $|s=4\}$ [the subspace up to generalized seniority $S = 2s = 8$, see Eq. (\ref{le_s_space})]. The Hamiltonian matrix in M-scheme has dimension $646,430$ (for $M=0$ including both parities). The basis (\ref{sen_basis}) of the space $|s=4\}$ is in general not orthogonal, and the non-trivial overlap matrix of the basis has the same dimension. By the recent fast algorithm of generalized seniority \cite{Jia_2015}, we compute the Hamiltonian and the overlap matrices that are both {\emph{sparse}}. The generalized (nonorthogonal basis) eigenvalue problem is solved by the Matlab function ``eigs'' in the Lanczos method for the lowest $5000$ eigenstates of each parity. In Ref. \cite{Jia_2016} we proved that the truncation up to an arbitrary generalized seniority preserves the particle-hole symmetry; the particle $|s\}$ and the hole $|s\}$ are the same subspace [$0 \le s \le \min(N,\Omega-N)$, where $2\Omega = \sum_j (2j + 1)$]. At a given nucleus the results are independent of choosing the particles or the holes as the degree of freedom. Practically we calculate $^{108 \sim 116}$Sn in the particle representation and $^{118 \sim 124}$Sn in the hole representation. The results for $^{108}$Sn and $^{124}$Sn are the exact shell-model ones because all the pairs (of particles or holes) are broken; others are approximations.

\begin{figure}
\includegraphics[width = 0.45\textwidth]{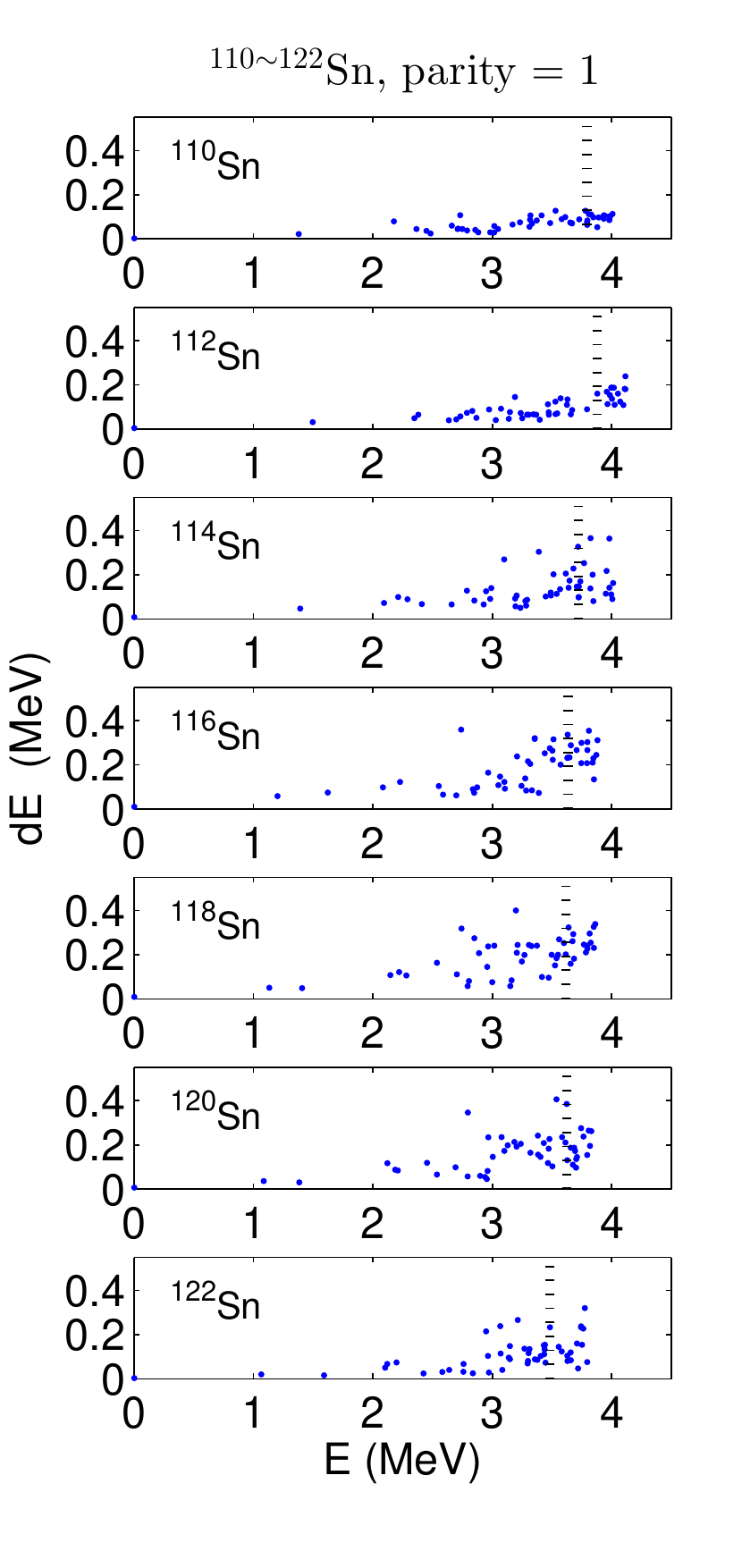}
\caption{\label{Fig_E_error_P0} (Color online) Errors of the generalized-seniority eigen energies for the lowest fifty positive-parity eigenstates in $^{110 \sim 122}$Sn. Every panel has fifty data points, each represents one $J$-state. The horizontal coordinate is the exact excitation energy from the shell-model calculation. The vertical coordinate is the error of the eigen energy from the generalized-seniority calculation, relative to the exact shell-model eigen energy. The vertical dotted line is the ``Fermi surface'' at the dimension of $s=1$. }
\end{figure}

\begin{figure}
\includegraphics[width = 0.45\textwidth]{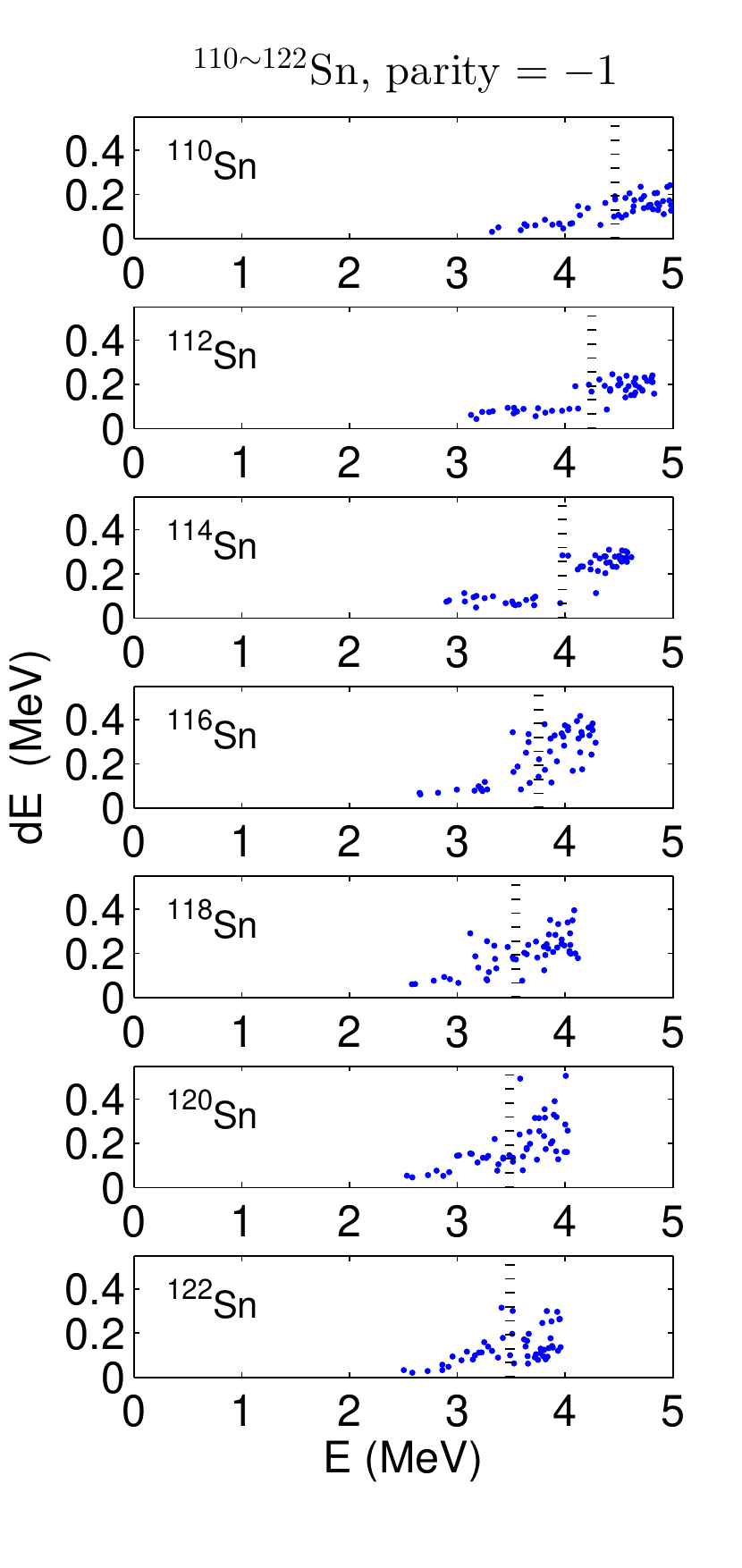}
\caption{\label{Fig_E_error_P1} (Color online)  Errors of the generalized-seniority eigen energies for the lowest fifty negative-parity eigenstates in $^{110 \sim 122}$Sn.  }
\end{figure}

To evaluate the quality of the approximation, we perform large-scale shell-model calculations for the lowest fifty eigenstates of each parity in $^{110 \sim 122}$Sn. Figures \ref{Fig_E_error_P0} and \ref{Fig_E_error_P1} show the errors $d E$ of the generalized-seniority eigen energies, relative to the exact shell-model ones. The vertical dotted line on each panel represents the ``Fermi surface'' at the dimension of $s = 1$. To the left of this line the number of data points is equal to the dimension of the $|s=1\}$ subspace (\ref{le_s_space}). We see in Fig. \ref{Fig_E_error_P0} that the ground-state energies converge very well. The actual numbers are $dE = 3, 4, 7, 10, 9, 7, 3$ keV for $^{110 \sim 122}$Sn. Below the $s=1$ Fermi surface the errors from Figs. \ref{Fig_E_error_P0} and \ref{Fig_E_error_P1} are generally small running from around $10$ keV to about $100$ keV. In most cases the errors have a sudden increase beyond the $s=1$ Fermi surface, to $200 \sim 300$ keV. This is related to the breakup of the second condensed pair. The convergence for higher excited states by the generalized-seniority calculation will be demonstrated later in another way.

\begin{figure}
\includegraphics[width = 0.45\textwidth]{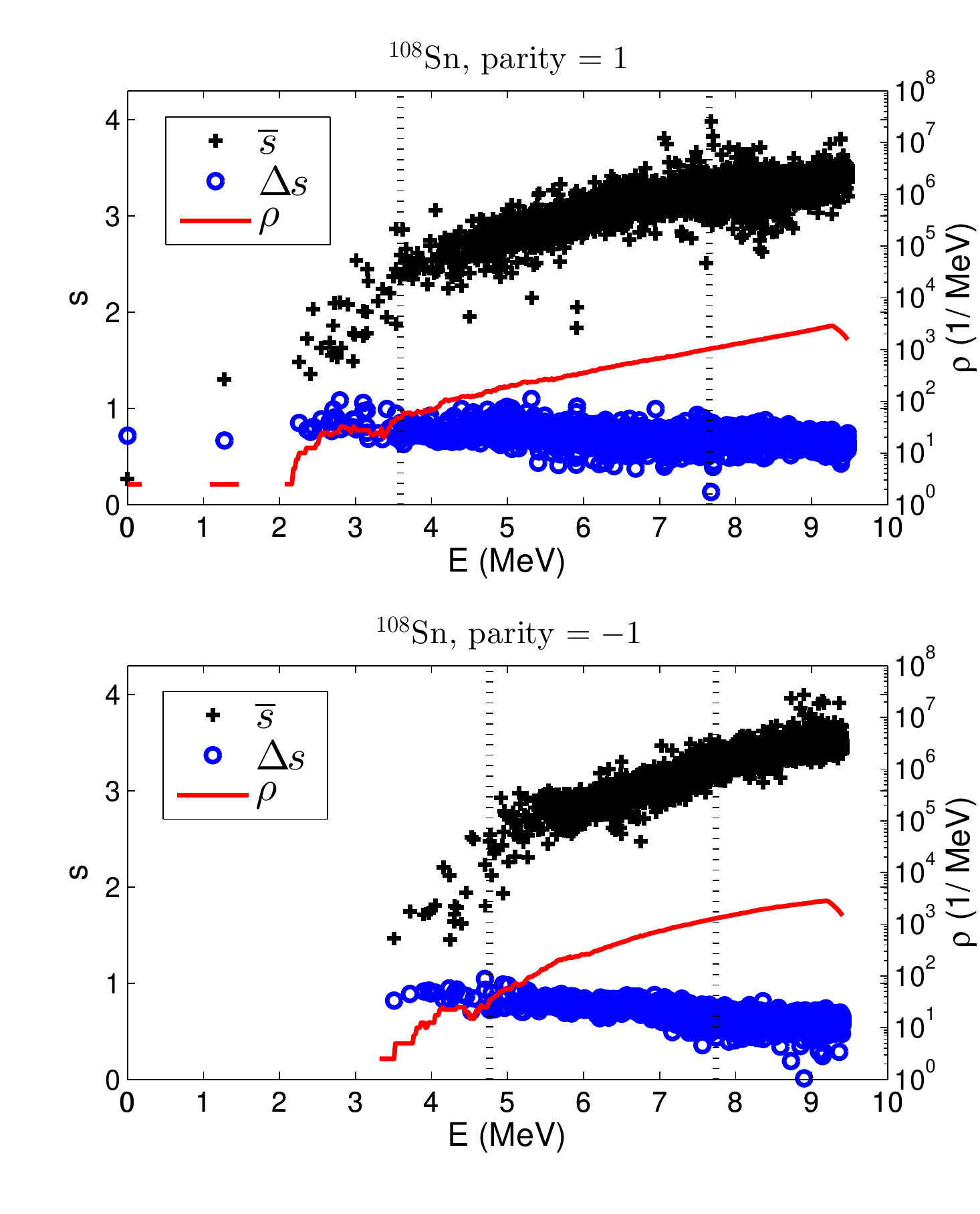}
\caption{\label{Fig_108_sbar} (Color online)  The generalized-seniority ($S=2s$) mean $\bar{s}$ and fluctuation $\Delta s$, and the $J$-level density $\rho$,
versus the excitation energy in $^{108}$Sn. The upper (lower) panel plots the lowest $5000$ eigenstates with positive (negative) parity. Each black plus (blue circle) symbol represents one state; its horizontal coordinate is the excitation energy, and the vertical coordinate is $\bar{s}$ ($\Delta s$) corresponding to the left axis. The red solid line corresponding to the right axis plots the $J$-level density $\rho$ averaged over a energy bin of $0.4$ MeV. The two vertical dotted lines are the ``Fermi surfaces'' at the dimension of $s = 1$ and $s = 2$, respectively. }
\end{figure}
\begin{figure}
\includegraphics[width = 0.45\textwidth]{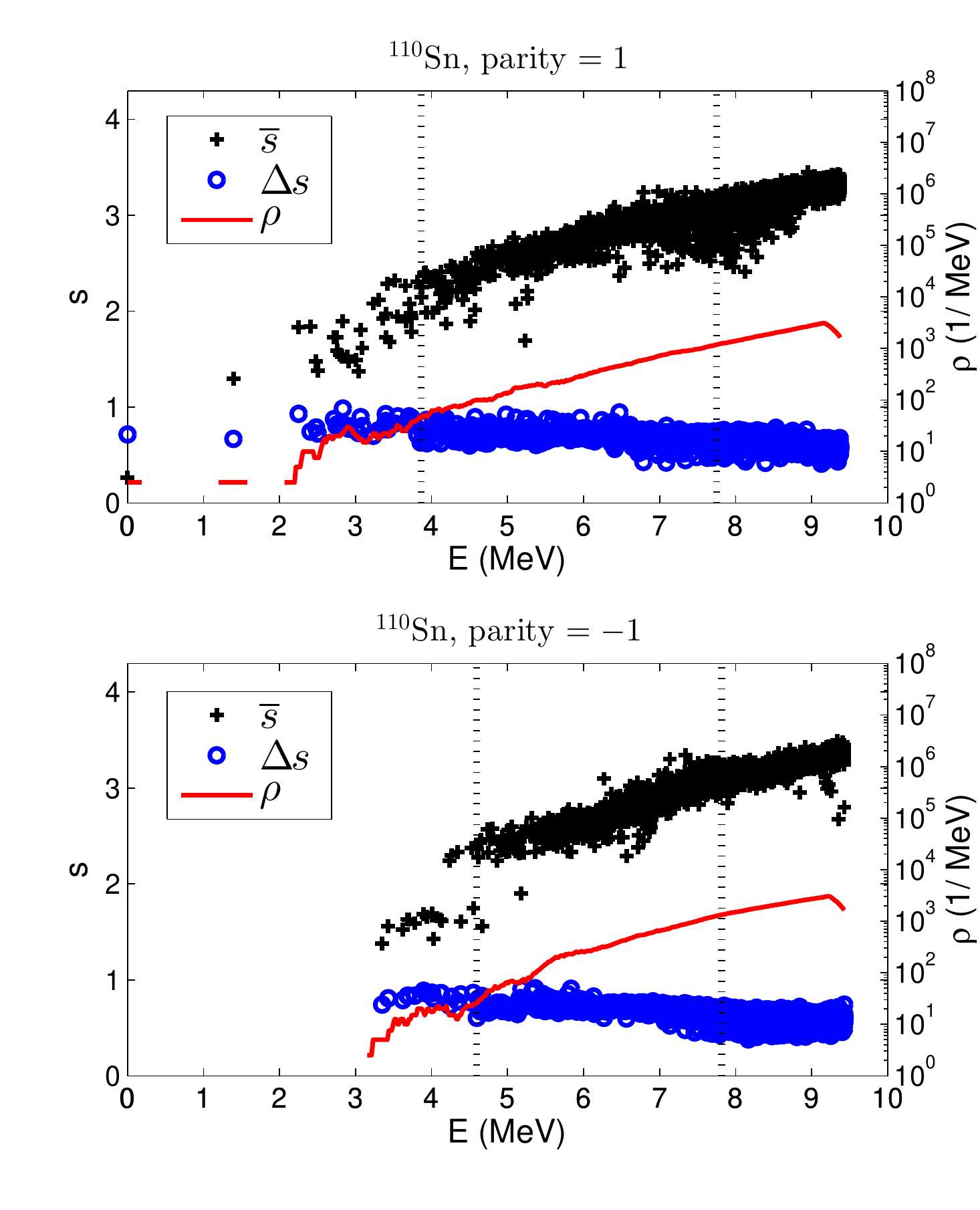}
\caption{\label{Fig_110_sbar} (Color online) The generalized-seniority mean and fluctuation, and the $J$-level density, versus the excitation energy in $^{110}$Sn.}
\end{figure}
\begin{figure}
\includegraphics[width = 0.45\textwidth]{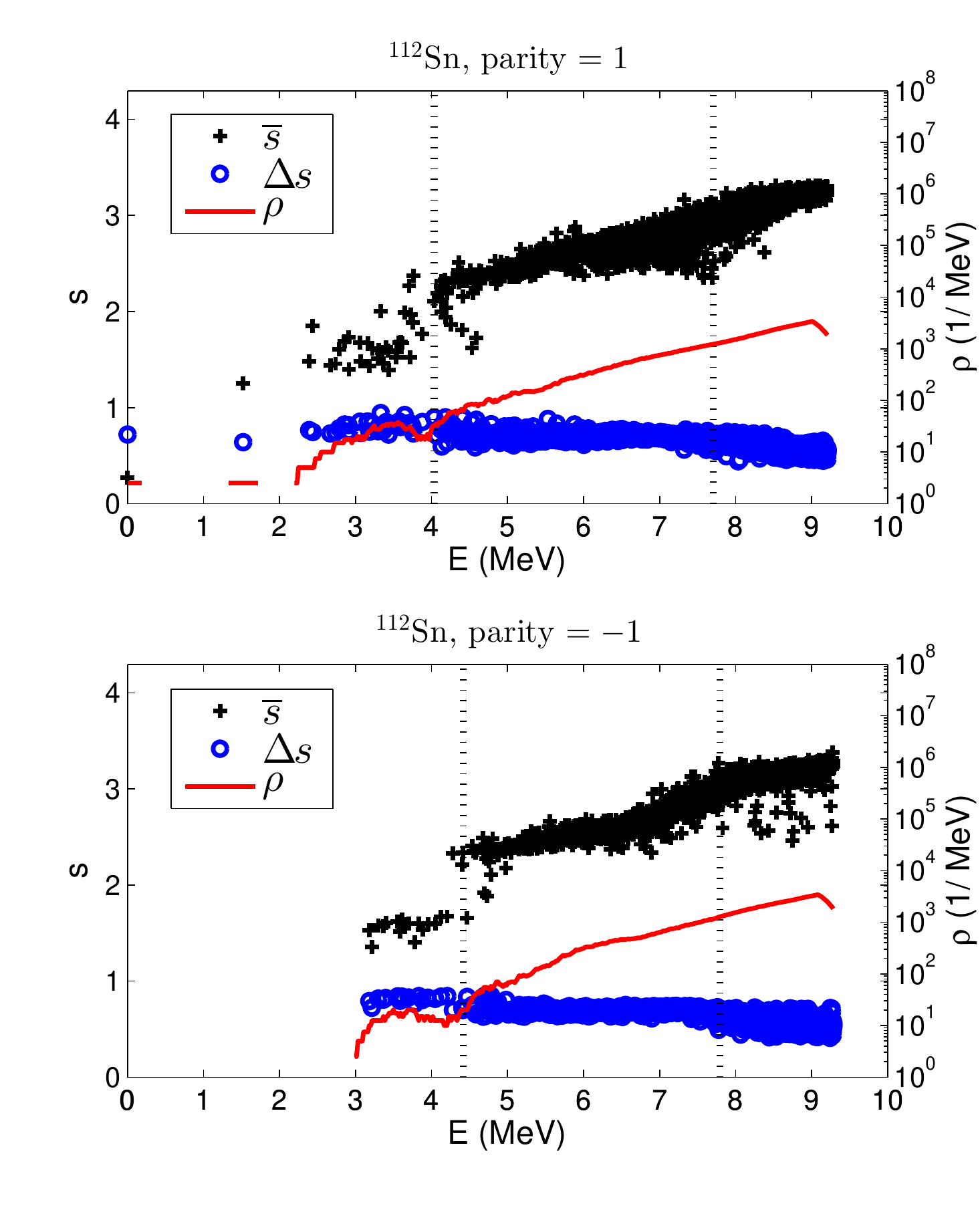}
\caption{\label{Fig_112_sbar} (Color online) The generalized-seniority mean and fluctuation, and the $J$-level density, versus the excitation energy in $^{112}$Sn. }
\end{figure}
\begin{figure}
\includegraphics[width = 0.45\textwidth]{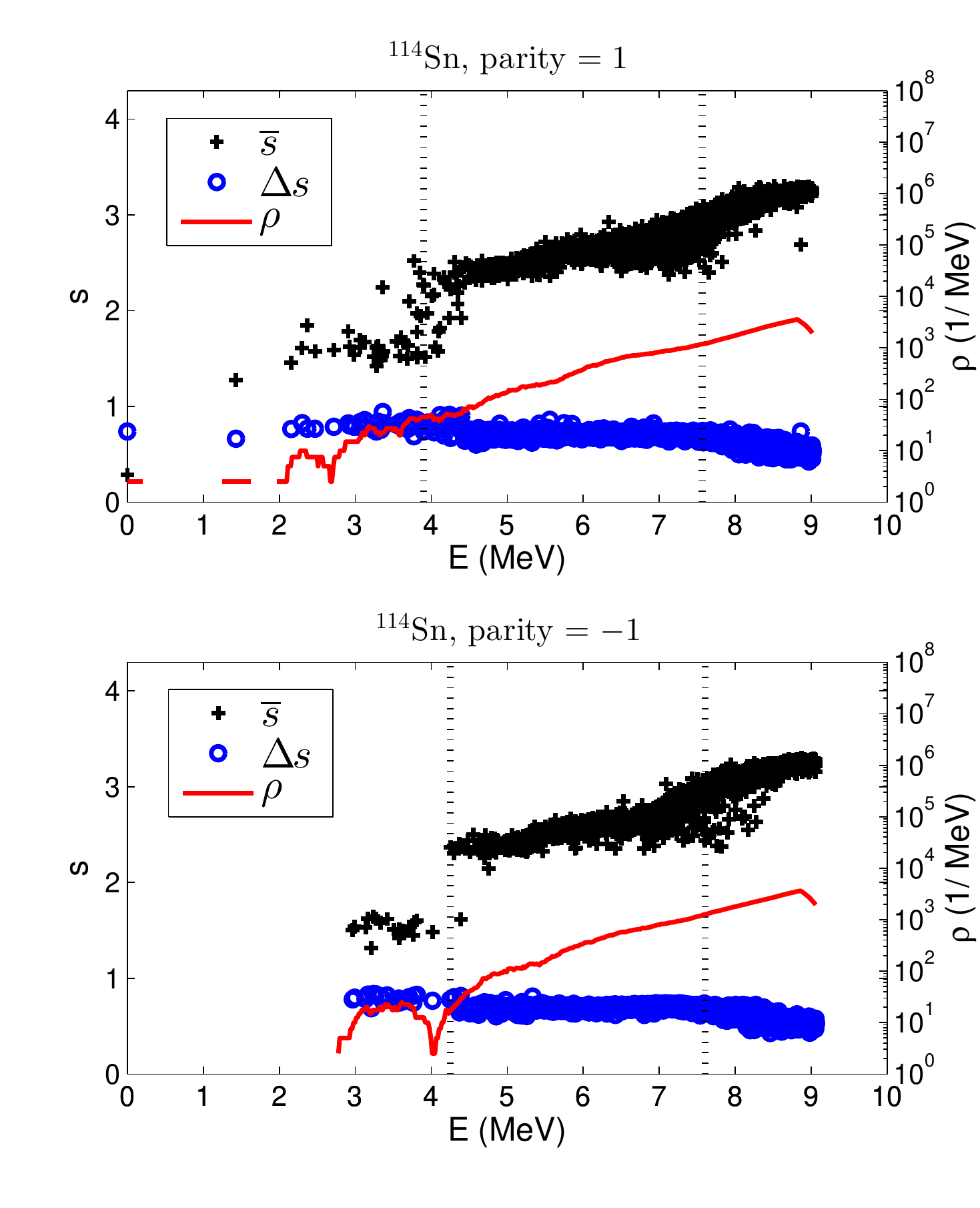}
\caption{\label{Fig_114_sbar} (Color online) The generalized-seniority mean and fluctuation, and the $J$-level density, versus the excitation energy in $^{114}$Sn. }
\end{figure}
\begin{figure}
\includegraphics[width = 0.45\textwidth]{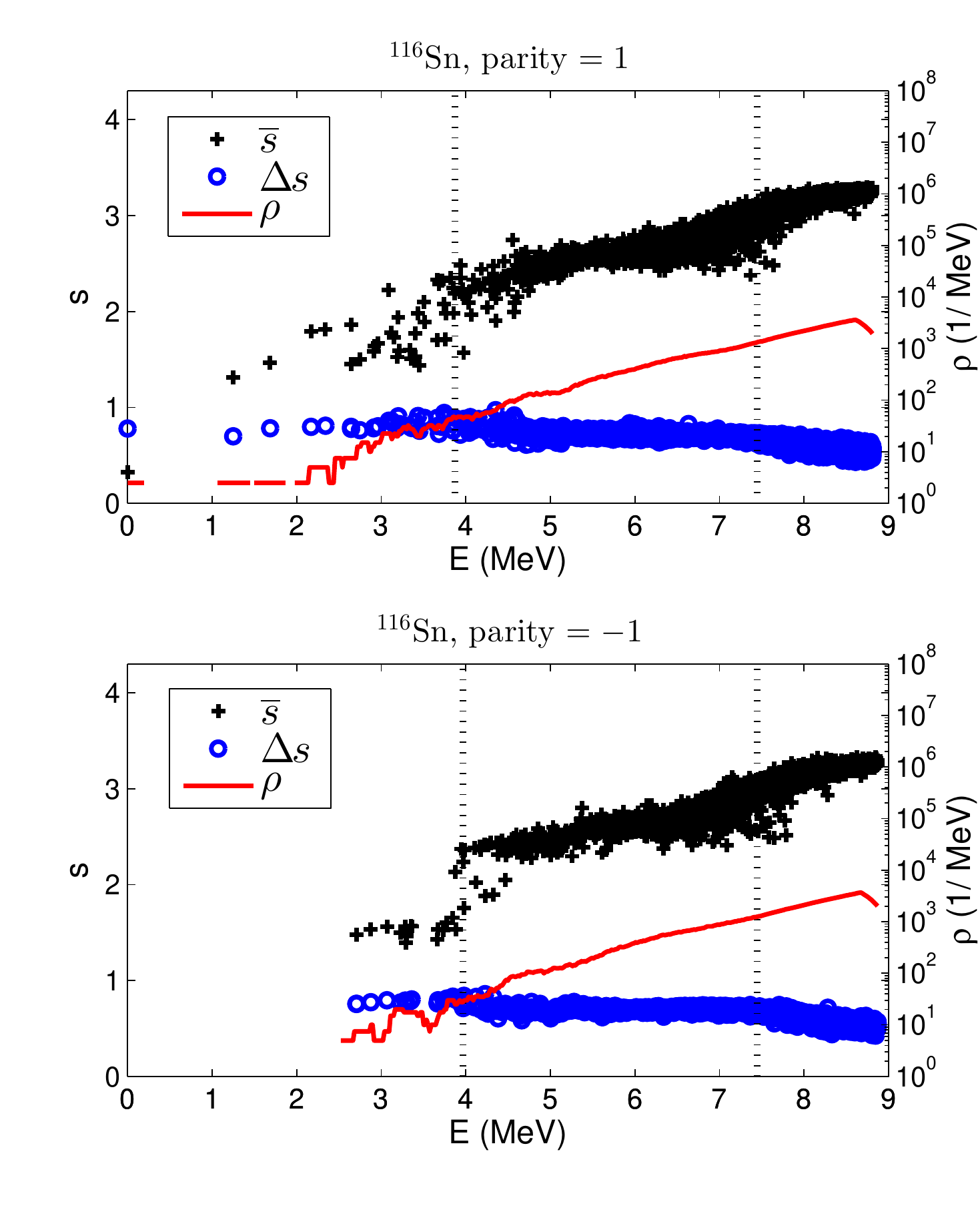}
\caption{\label{Fig_116_sbar} (Color online) The generalized-seniority mean and fluctuation, and the $J$-level density, versus the excitation energy in $^{116}$Sn. }
\end{figure}
\begin{figure}
\includegraphics[width = 0.45\textwidth]{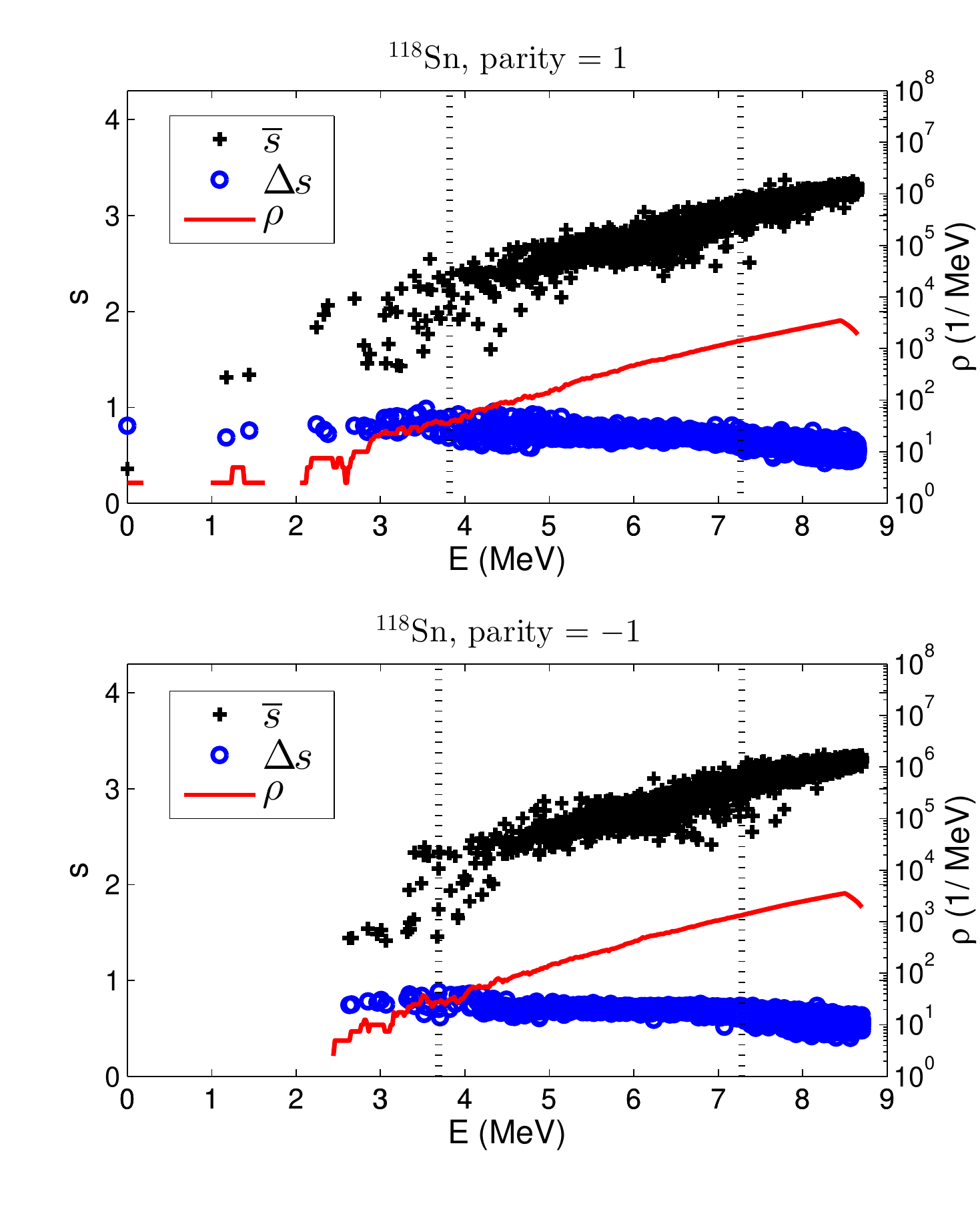}
\caption{\label{Fig_118_sbar} (Color online) The generalized-seniority mean and fluctuation, and the $J$-level density, versus the excitation energy in $^{118}$Sn. }
\end{figure}
\begin{figure}
\includegraphics[width = 0.45\textwidth]{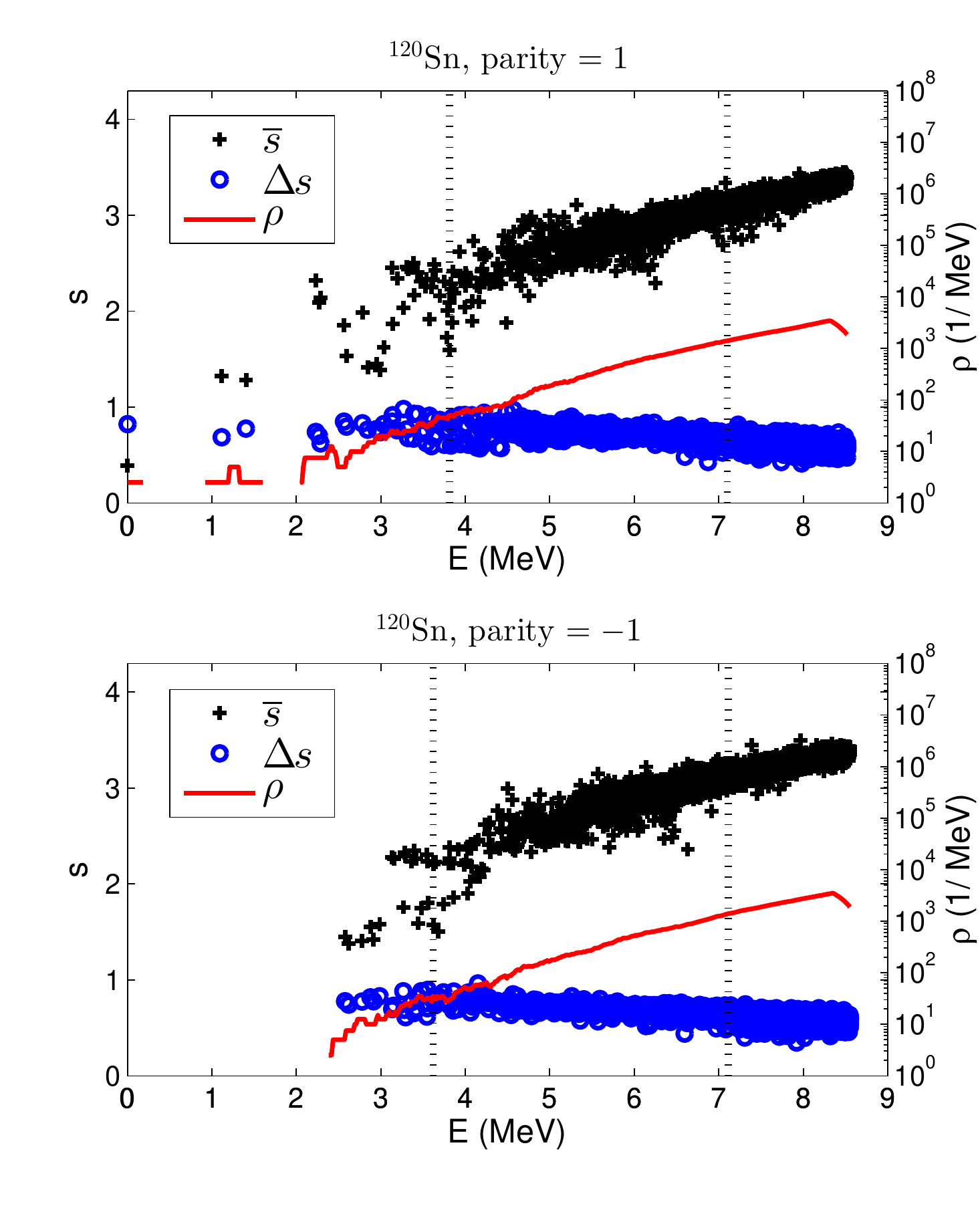}
\caption{\label{Fig_120_sbar} (Color online) The generalized-seniority mean and fluctuation, and the $J$-level density, versus the excitation energy in $^{120}$Sn. }
\end{figure}
\begin{figure}
\includegraphics[width = 0.45\textwidth]{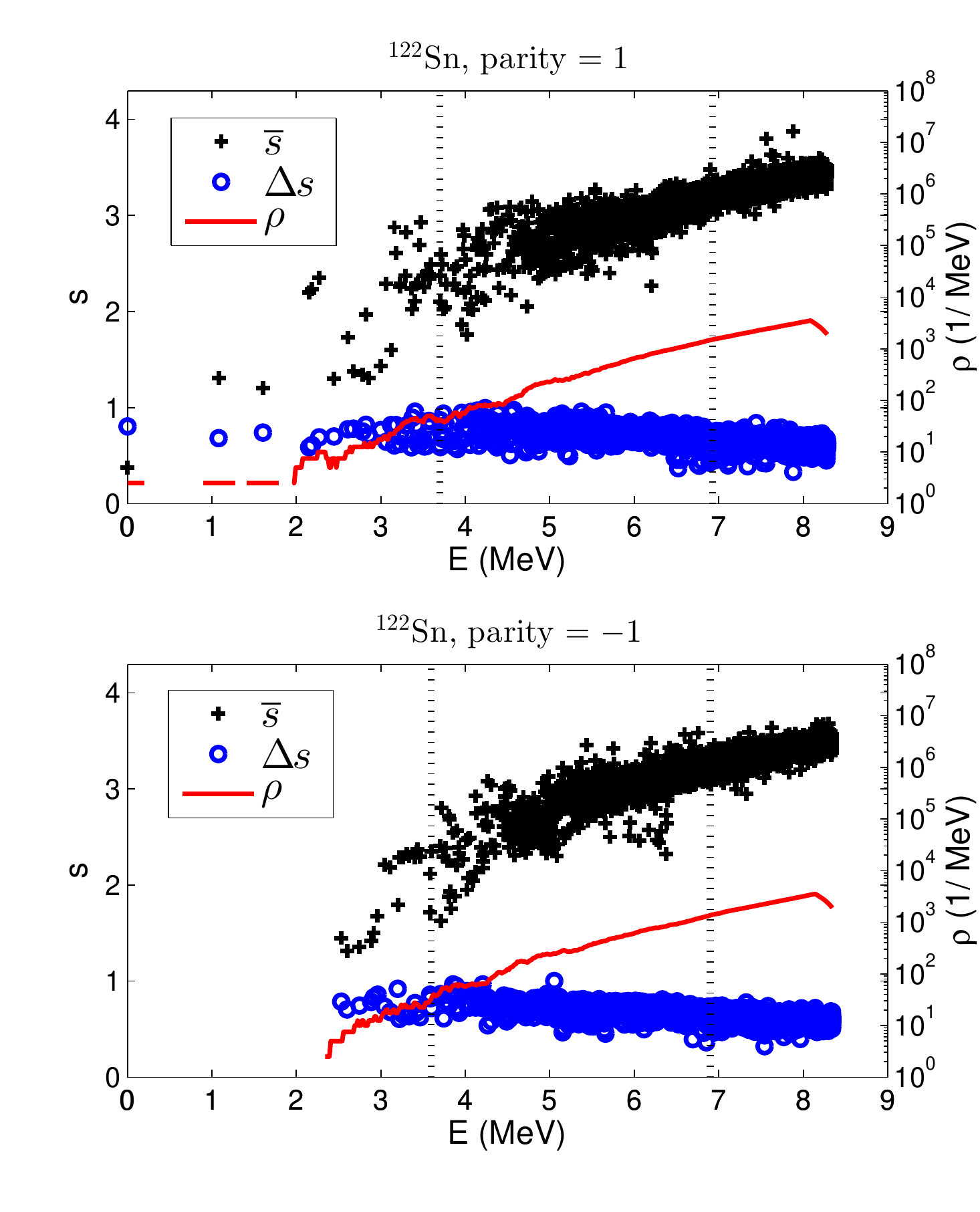}
\caption{\label{Fig_122_sbar} (Color online) The generalized-seniority mean and fluctuation, and the $J$-level density, versus the excitation energy in $^{122}$Sn. }
\end{figure}
\begin{figure}
\includegraphics[width = 0.45\textwidth]{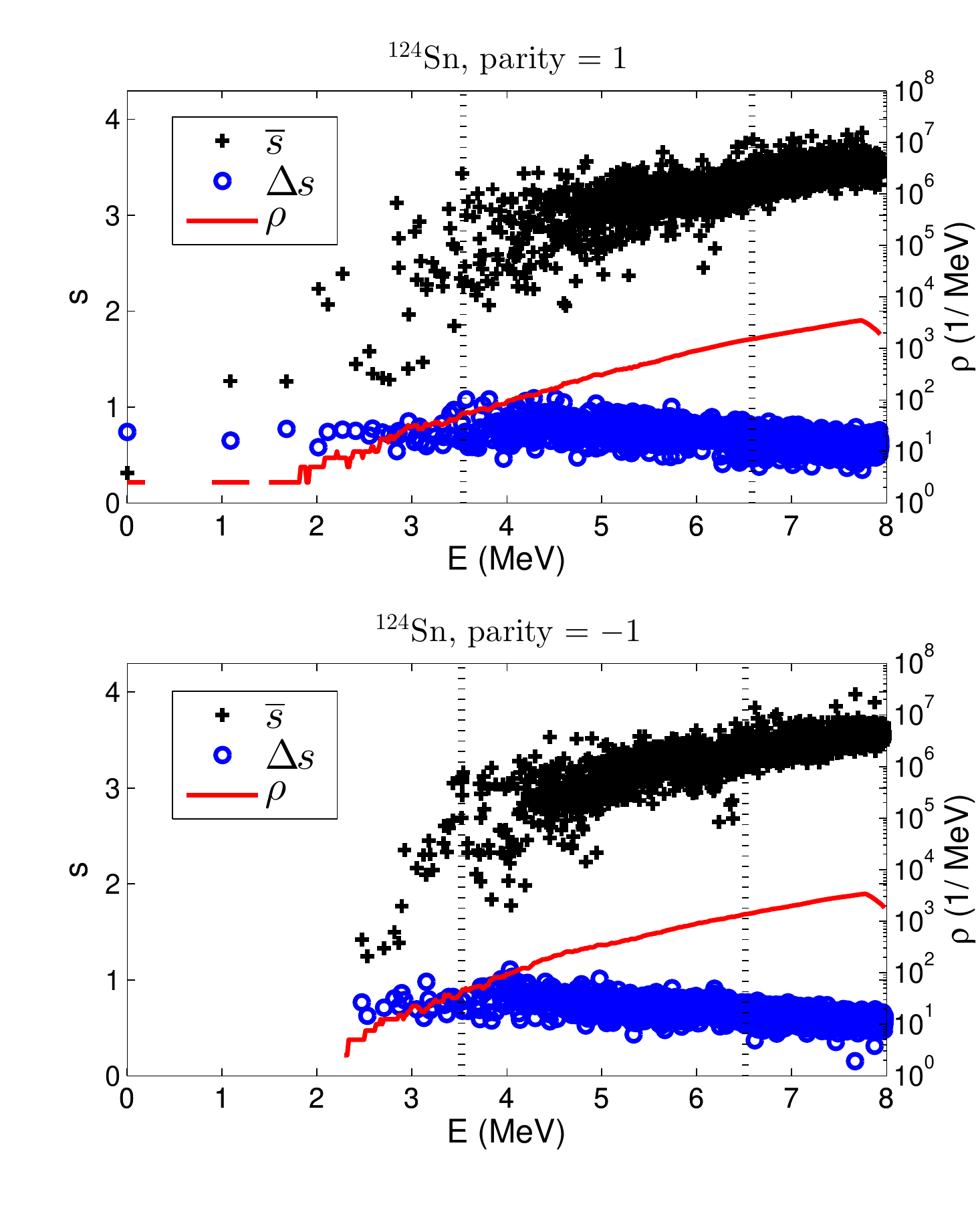}
\caption{\label{Fig_124_sbar} (Color online) The generalized-seniority mean and fluctuation, and the $J$-level density, versus the excitation energy in $^{124}$Sn. }
\end{figure}

For each eigen wavefunction, we compute its generalized-seniority amplitudes $P(s)$ (\ref{P_s}), mean $\bar{s}$ (\ref{s_bar}), and fluctuation $\Delta s$ (\ref{Delta_s}). The results of $\bar{s}$ and $\Delta s$ are plotted in Figs. \ref{Fig_108_sbar}-\ref{Fig_124_sbar}, together with the $J$-level density
\begin{eqnarray}
\rho(E) \equiv \frac{ \Omega (E-dE,E+dE) } { 2dE } ,  \label{rho_E}
\end{eqnarray}
where $\Omega (E-dE,E+dE)$ is the number of $J$-levels in the energy interval $(E-dE,E+dE)$, and $dE = 0.2$ MeV. The drop of the $\rho(E)$ curves at the large-energy end is artificial and simply because the energy cutoff ($5000$ Lanczos states) is reached, beyond which the $J$-levels are not computed thus absent from Eq. (\ref{rho_E}).


Intuitively, from pure pairing models we would expect a relatively sharp staircase curve of $\bar{s}$; breaking each pair costs roughly the pairing energy $2\Delta$ (in BCS language). In Figs. \ref{Fig_108_sbar}-\ref{Fig_124_sbar} the realistic $\bar{s}$ curves indeed show such a staircase pattern, although blurred by generalized-seniority mixing interactions. In general the staircase pattern is more pronounced near the mid-shell. This is consistent with the conventional wisdom that in the exact seniority scheme the collective pairing effect is proportional to $N(\Omega-N)$. Usually the mid-shell region has the largest effective model space ($\Omega_{\rm{eff}}$) available for the collective pairing; the Fermi surface sits at the middle, both upper and lower single-particle levels participate. This region also has the largest number of pairs $N \lesssim \Omega/2$ (particle and hole representations in the lower and upper shell, respectively). Therefore near the mid-shell the collective pairing effect is the most enhanced and the pair condensate is the best developed. As a signature the staircase curve is the sharpest. Figures \ref{Fig_112_sbar}-\ref{Fig_118_sbar} show that in mid-shell nuclei $^{112 \sim 118}$Sn the pairing force dominates over other correlations and the superfluid structure persists at higher energies in the increasingly dense spectrum. The two vertical dotted lines on each figure represent the ``Fermi surfaces'' at the dimension of $s = 1$ and
$s = 2$. To the left of the first (second) line the number of data points is equal to the dimension of the $|s=1\}$ ($|s=2\}$) subspace (\ref{le_s_space}). We see that the transitions in the $\bar{s}$ curves happen roughly at the Fermi surfaces, especially so for $s=1$. In this sense the generalized seniority should be the ultimate truncation scheme for the full low-lying spectrum in mid-shell tin isotopes; no further dimension reduction is possible. The $s=2$ Fermi surface is slightly below twice the $s=1$ Fermi surface. It is consistent with the exact seniority scheme where this ratio is $2(1-1/\Omega)$. If we take $\Omega = 16$, this ratio is $1.875$, quite consistent with those in Figs. \ref{Fig_112_sbar}-\ref{Fig_118_sbar}.

On the other hand, Figs. \ref{Fig_108_sbar}, \ref{Fig_122_sbar}, and \ref{Fig_124_sbar} show that away from the mid-shell the staircase pattern is less obvious in $^{108, 122, 124}$Sn. Here the available effective model space $\Omega_{\rm{eff}}$ is relatively small. The pair condensate is not well developed with the limited number of collective pairs. Other correlations may easily destroy the superfluid structure at higher energies.

Looking more closely, $^{114}$Sn has the sharpest staircase curve instead of the exact mid-shell nucleus $^{116}$Sn. This is consistent with the conclusion of Ref. \cite{Morales_2011}. The experimental $B(E2)$ values \cite{Banu_2005, Vaman_2007, Ekstrom_2008, Jungclaus_2011}, proportional to $N(\Omega-N)$ in the exact seniority scheme, in fact have a small dip at $^{116}$Sn in the generally parabolic [$N(\Omega-N)$] curve for tin isotopes. The dip indicates slightly reduced collective pairing effect, and was explained in Ref. \cite{Morales_2011} by the different filling rates of the two groups of $j$-orbits. The lower group has two orbits $0g_{7/2}$ and $1d_{5/2}$, and the higher group has three orbits $2s_{1/2}$, $1d_{3/2}$, and $0h_{11/2}$. The two groups separate by a moderate energy difference. In the lower shell mainly the lower group contributes to the coherent pairing, while in the upper shell mainly the higher group contributes. At $^{116}$Sn the occupation is switching between the two groups and results in slightly reduced pairing effect. Another observation is that the pattern is sharper for negative-parity states than that for positive-parity states. Thus the generalized-seniority mixing matrix elements are smaller involving the intruder orbit $0h_{11/2}$. We also notice that in Figs. \ref{Fig_108_sbar}-\ref{Fig_124_sbar} some data points have large $\bar{s}$ but small excitation energy $E$ (lying toward the top-left corner). These states should be identified as collective states and deserve more attention.

Figures \ref{Fig_108_sbar}-\ref{Fig_124_sbar} show that in general the $J$-level density in a logarithmic scale increases linearly with the excitation energy. But in $^{108 \sim 114}$Sn the curves show an apparent dip around the $s=1$ Fermi surface. This is another piece of evidence for the persisting superfluid structure; the $s = 1$ broken-pair states have close energies and this group is quite separated in energy from the group of $s = 2$ broken-pair states.

The $J$-level densities up to the neutron separation energy in $^{116,118,122}$Sn have been extracted experimentally under certain assumptions \cite{Agvaanluvsan_2009, Toft_2010, Toft_2011}. However the computed $J$-level densities of this work are smaller than the experimentally extracted ones near the neutron separation energy. This is not surprising, because at high energies the cross-shell excitations become important but these are missing in the current model space. In this work we limit the valence space to the neutron $50 \sim 82$ major shell. We also notice from Figs. \ref{Fig_E_error_P0} and \ref{Fig_E_error_P1} that systematically the generalized-seniority energies lie slightly higher than the exact ones. Shifting the $J$-level density curve $\rho(E)$ to smaller energy (to the left) leads to better agreements with experiments. Another possibility is that the adopted interaction \cite{Qi_2012}, which was fitted by only the first few low-lying states, may not be very accurate near the neutron separation energy.

Above we see evidence for the persisting superfluid structure. Meanwhile, sizable generalized-seniority mixing exists in all the nuclei $^{108 \sim 124}$Sn, even near the mid-shell.  In Figs. \ref{Fig_108_sbar}-\ref{Fig_124_sbar}, the stairs of the $\bar{s}$ staircase curves never sit at integers; the first and the second stairs are close to $\bar{s} = 1.5$ and $2.5$, instead of $1$ and $2$. The fluctuation $\Delta s$ (\ref{Delta_s}) measures directly the degree of generalized-seniority mixing. In these figures $\Delta s$ is sizable around $0.7$ and decreases slightly with increasing excitation energy. At a given excitation energy, the spread of $\Delta s$ is rather small, hence nearby eigenstates have quite similar degree of generalized-seniority mixing. Practically no pure ($\Delta s \approx 0$) generalized-seniority state exists; the only three exceptions are a positive-parity state ($\sim 7.7$ MeV) and a negative one ($\sim 9$ MeV) in $^{108}$Sn, and a negative one in $^{124}$Sn ($\sim 7.7$ MeV). All the three states are at high excitation energies and appear in nuclei far from the mid-shell. The results suggest more care to schematic studies using pure generalized-seniority states. For example, the recent paper \cite{Maheshwari_2016} treated the isomer states $10^+$, $13^-$, $15^-$ in $^{116 \sim 130}$Sn as pure generalized-seniority states with $S = 2$, $4$, $4$, respectively. While the treatment was simple and inspiring, these isomer states in fact have mixed generalized-seniority based on the results of the current work. Similar comments apply to Ref. \cite{Morales_2011} that treated the first $2^+$ state as a pure $S = 2$ state.

\begin{figure}
\includegraphics[width = 0.45\textwidth]{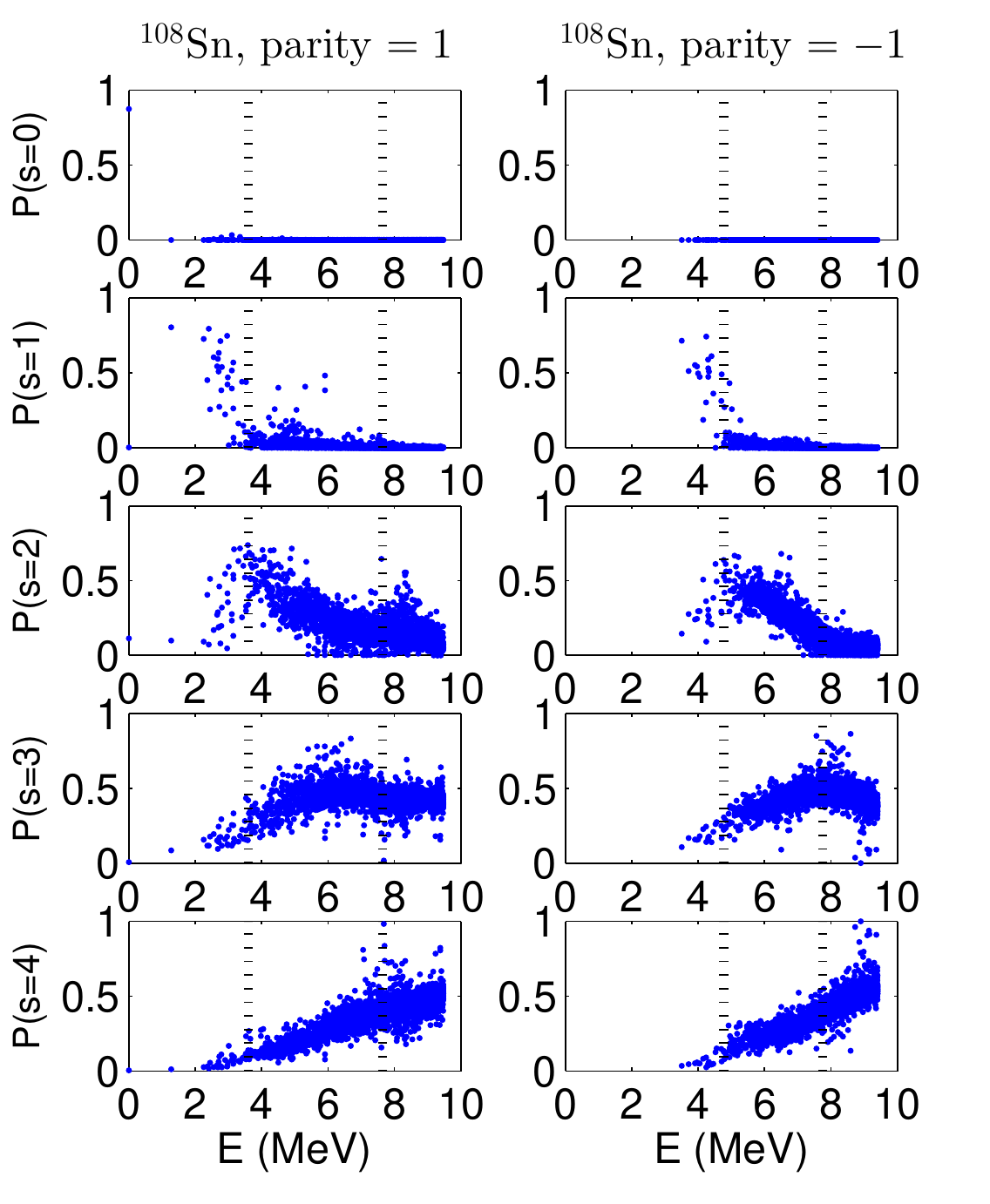}
\caption{\label{Fig_108_Ps} (Color online)  Amplitudes $P(s)$ of each generalized seniority $S=2s$ versus the excitation energy in $^{108}$Sn. The left (right) panels plot the lowest $5000$ eigenstates with positive (negative) parity. Therefore each panel has $5000$ data points. The two vertical dotted lines are the $s=1$ and $s=2$ ``Fermi surfaces'' at the same positions as those in Fig. \ref{Fig_108_sbar}. }
\end{figure}

\begin{figure}
\includegraphics[width = 0.45\textwidth]{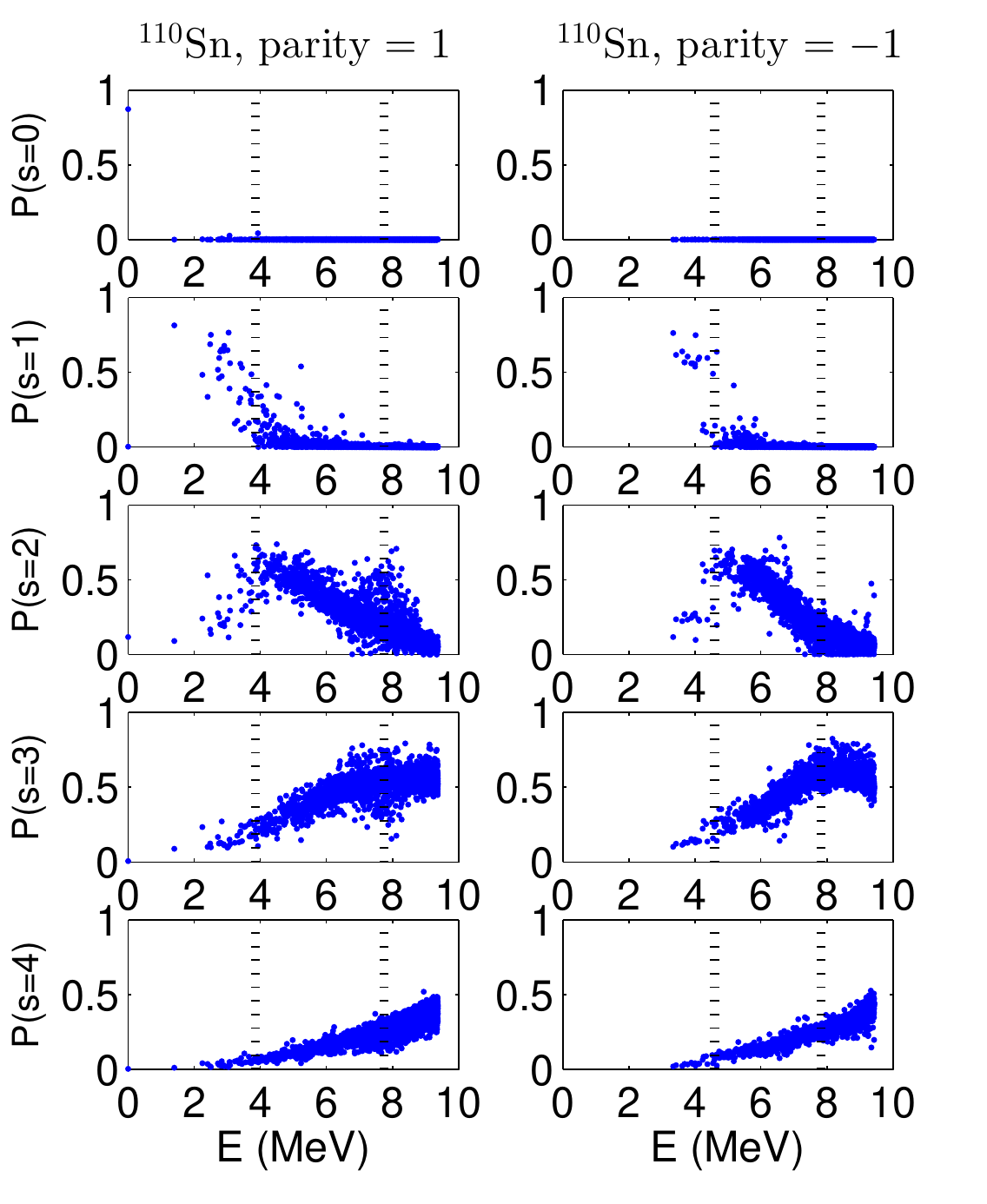}
\caption{\label{Fig_110_Ps} (Color online) Amplitudes of each generalized seniority versus the excitation energy in $^{110}$Sn. }
\end{figure}

\begin{figure}
\includegraphics[width = 0.45\textwidth]{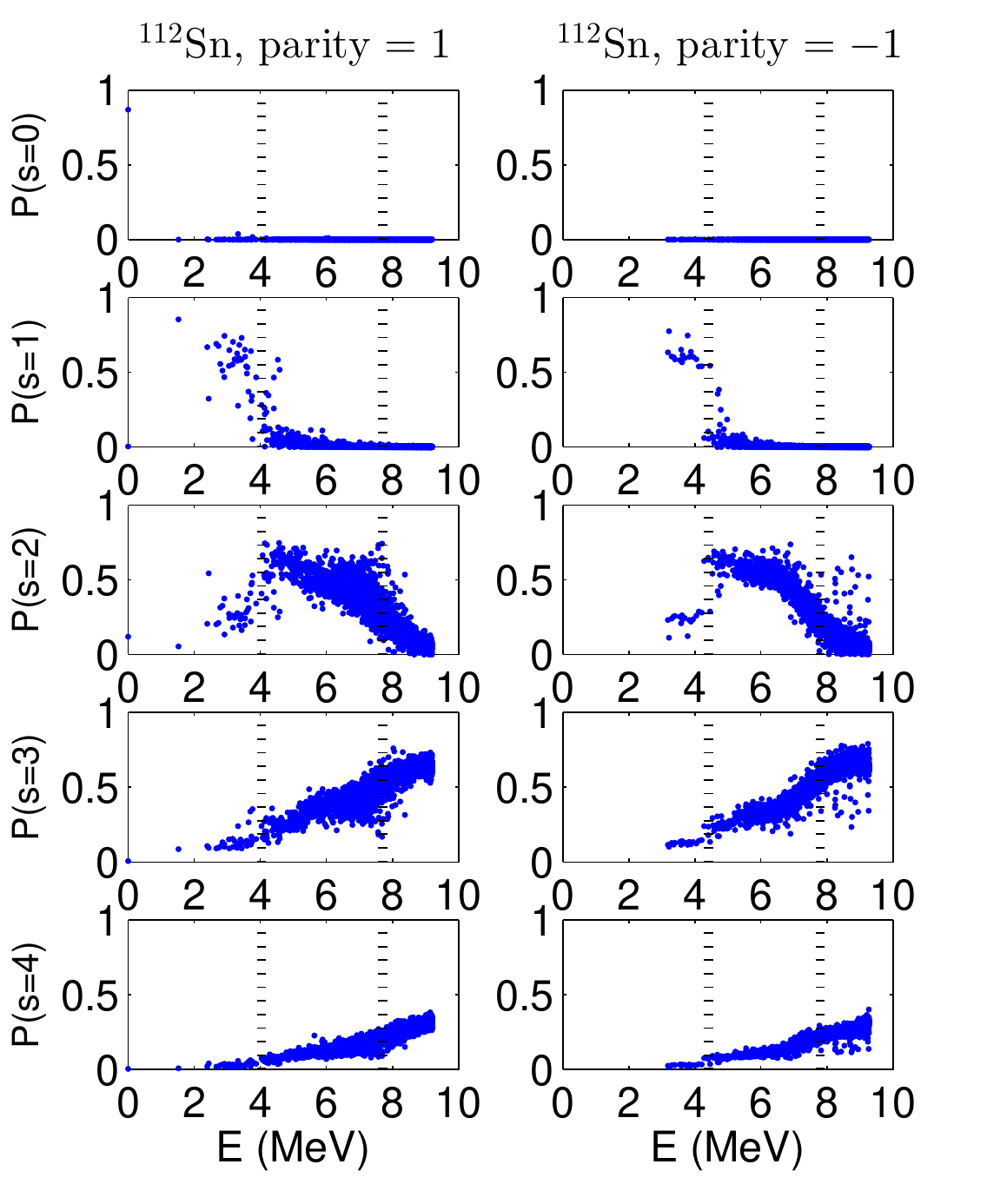}
\caption{\label{Fig_112_Ps} (Color online) Amplitudes of each generalized seniority versus the excitation energy in $^{112}$Sn. }
\end{figure}

\begin{figure}
\includegraphics[width = 0.45\textwidth]{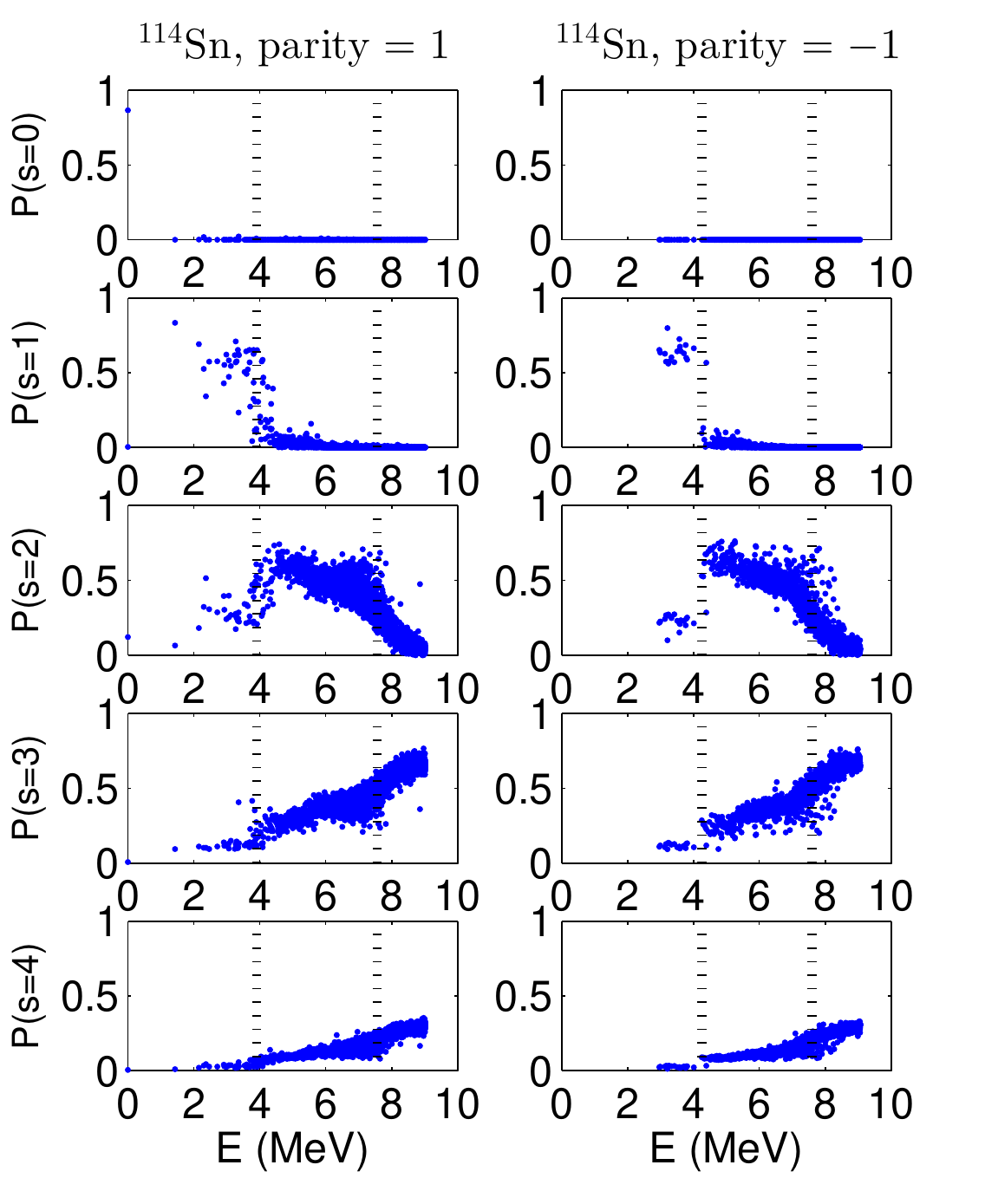}
\caption{\label{Fig_114_Ps} (Color online) Amplitudes of each generalized seniority versus the excitation energy in $^{114}$Sn. }
\end{figure}

\begin{figure}
\includegraphics[width = 0.45\textwidth]{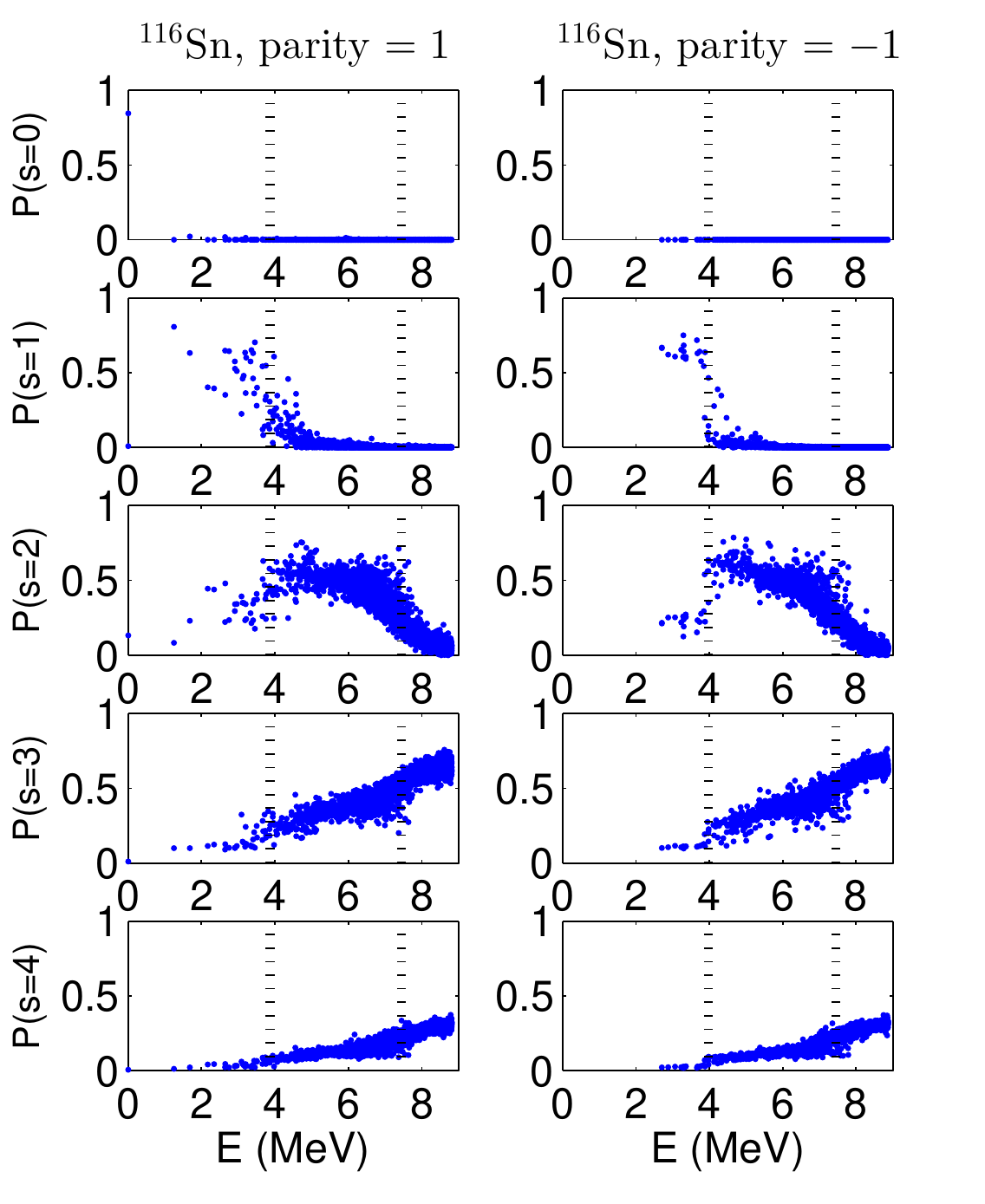}
\caption{\label{Fig_116_Ps} (Color online) Amplitudes of each generalized seniority versus the excitation energy in $^{116}$Sn. }
\end{figure}

\begin{figure}
\includegraphics[width = 0.45\textwidth]{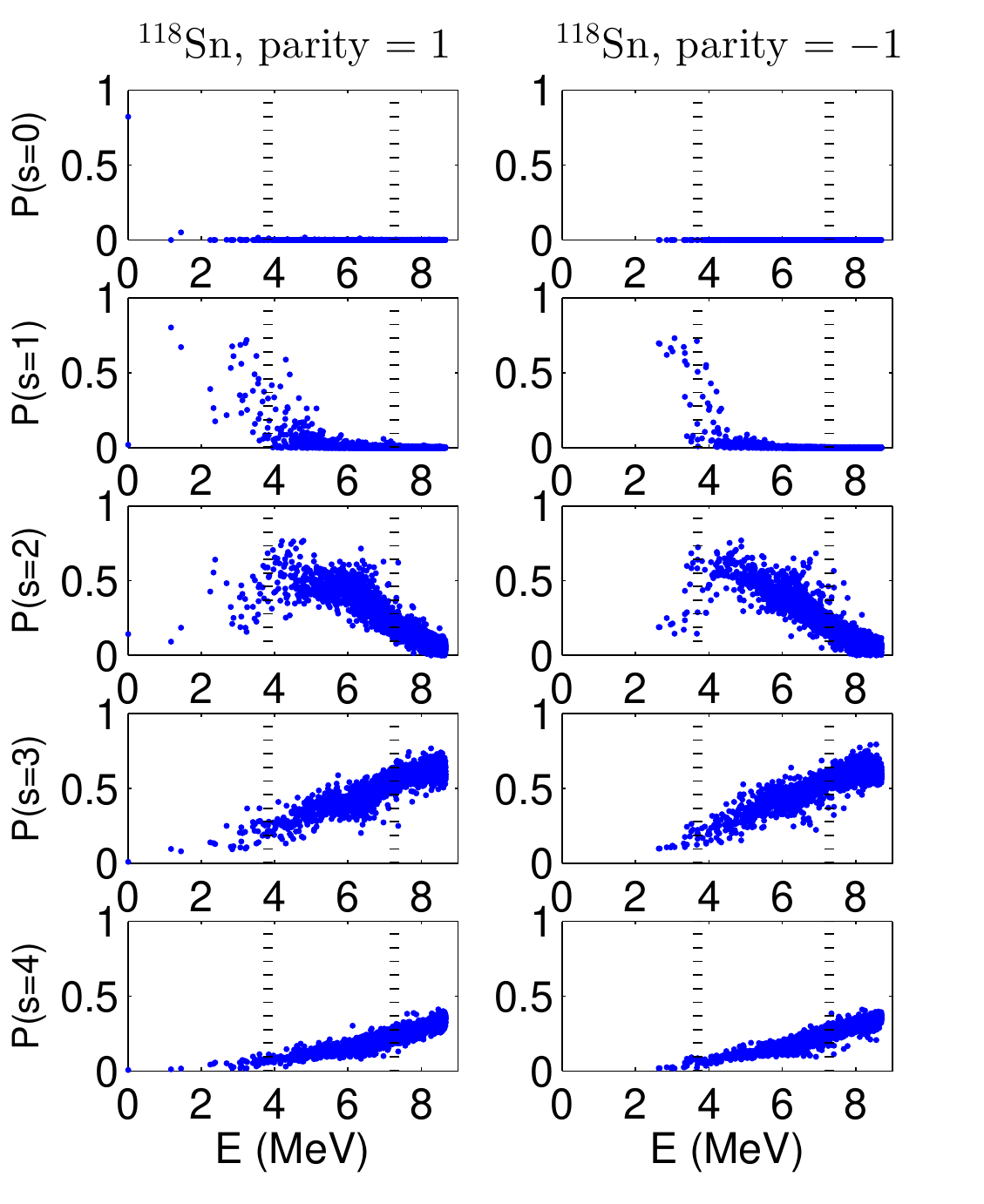}
\caption{\label{Fig_118_Ps} (Color online) Amplitudes of each generalized seniority versus the excitation energy in $^{118}$Sn. }
\end{figure}

\begin{figure}
\includegraphics[width = 0.45\textwidth]{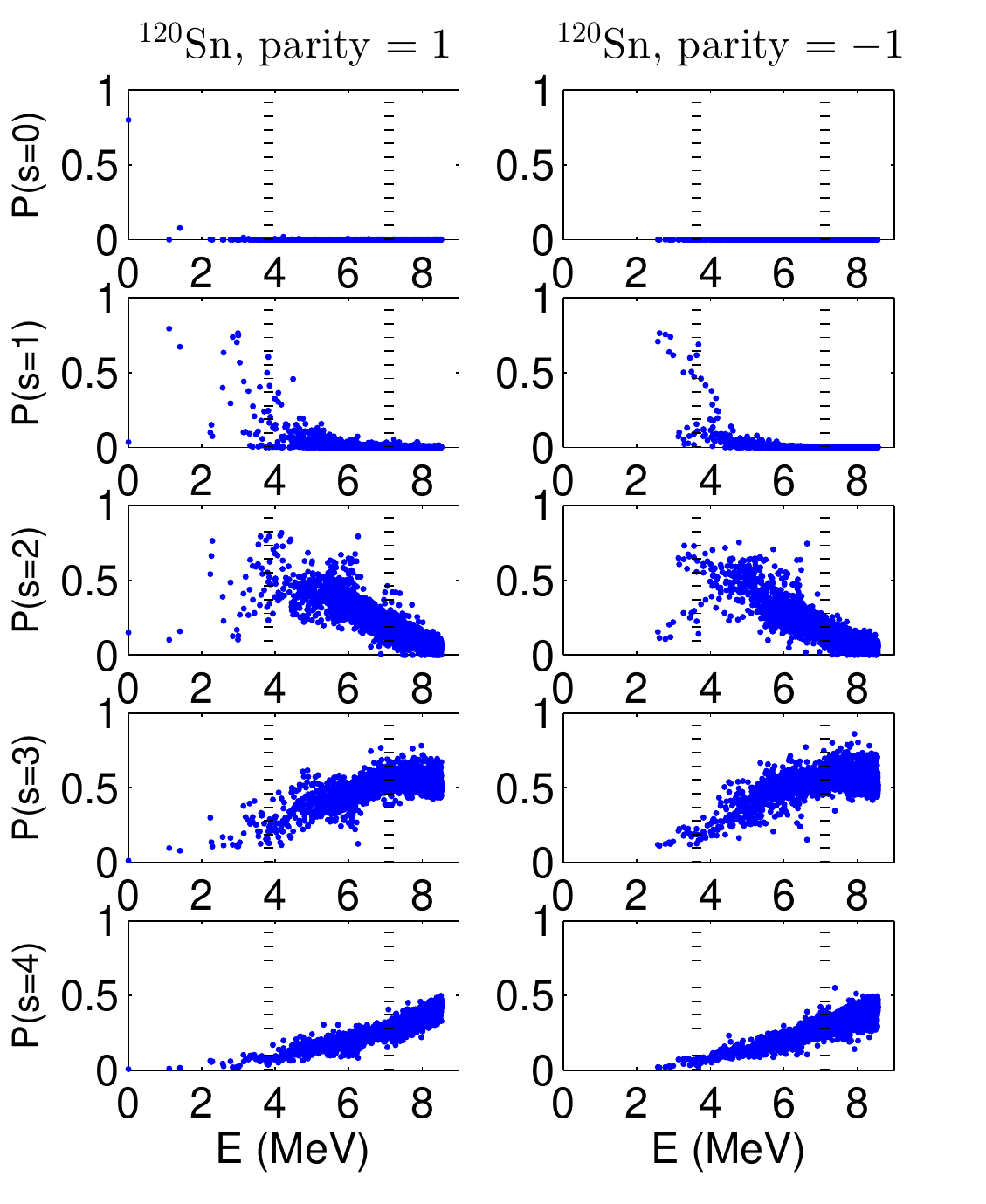}
\caption{\label{Fig_120_Ps} (Color online) Amplitudes of each generalized seniority versus the excitation energy in $^{120}$Sn. }
\end{figure}

\begin{figure}
\includegraphics[width = 0.45\textwidth]{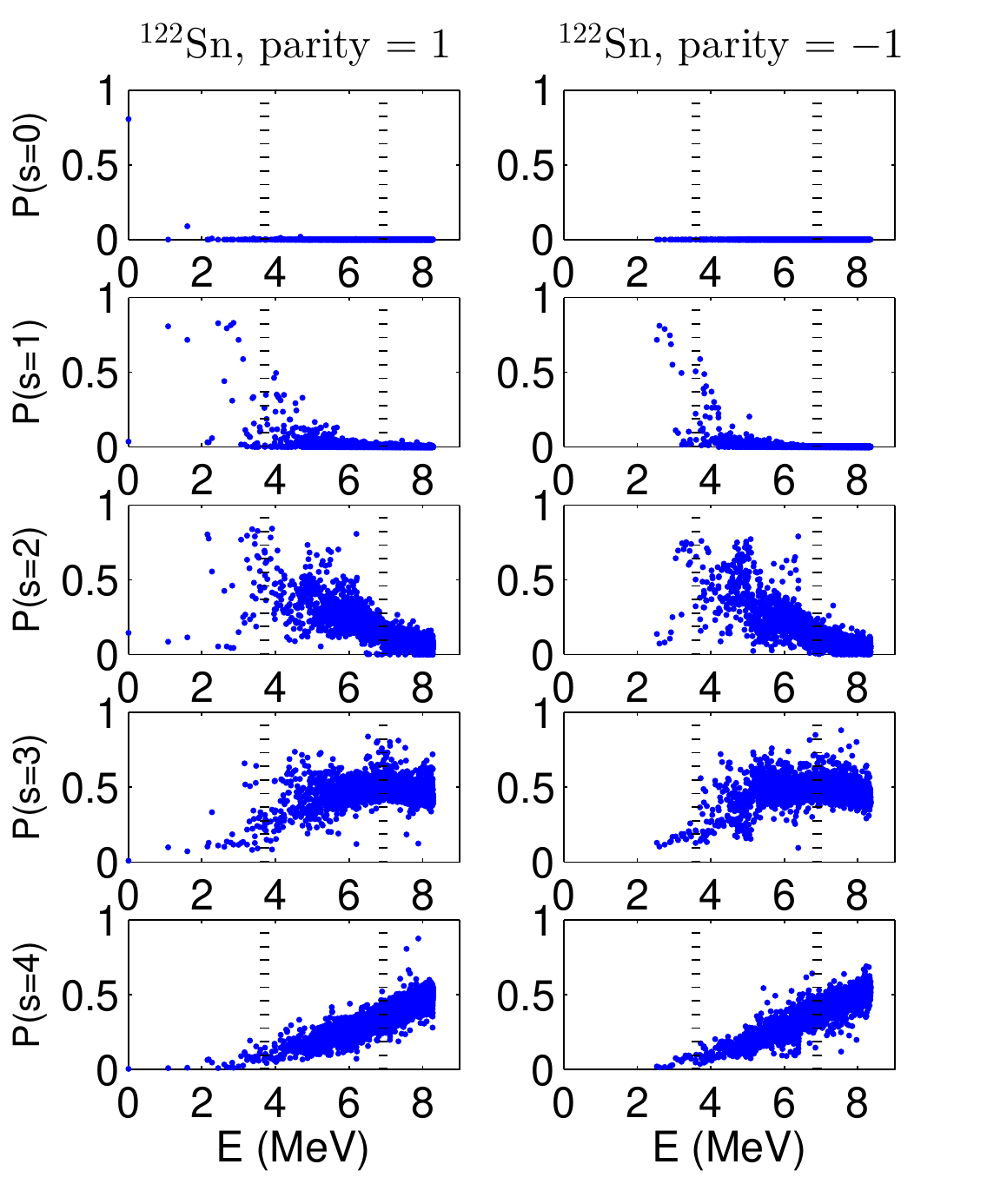}
\caption{\label{Fig_122_Ps} (Color online) Amplitudes of each generalized seniority versus the excitation energy in $^{122}$Sn. }
\end{figure}

\begin{figure}
\includegraphics[width = 0.45\textwidth]{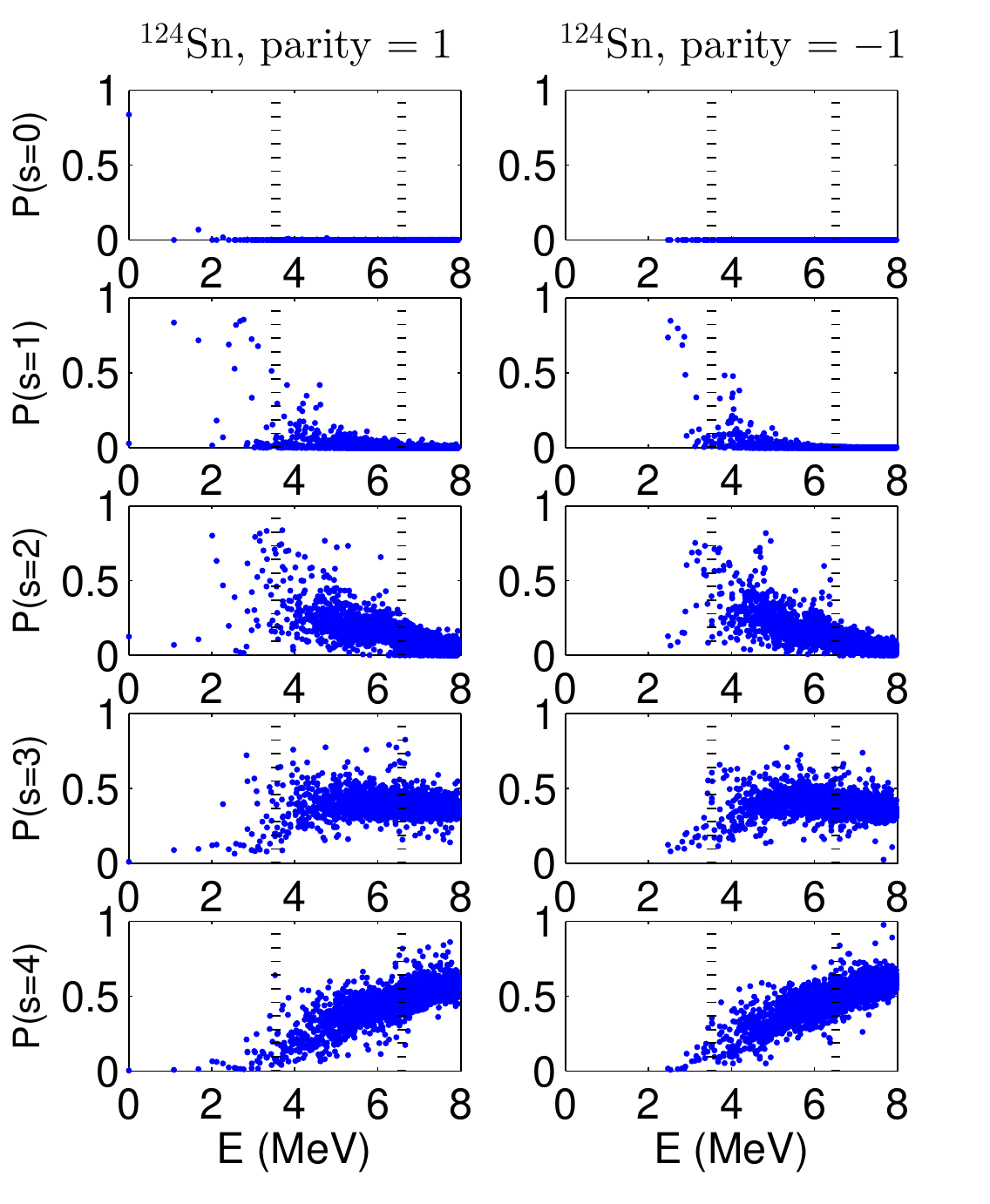}
\caption{\label{Fig_124_Ps} (Color online) Amplitudes of each generalized seniority versus the excitation energy in $^{124}$Sn. }
\end{figure}

\begin{figure}
\includegraphics[width = 0.45\textwidth]{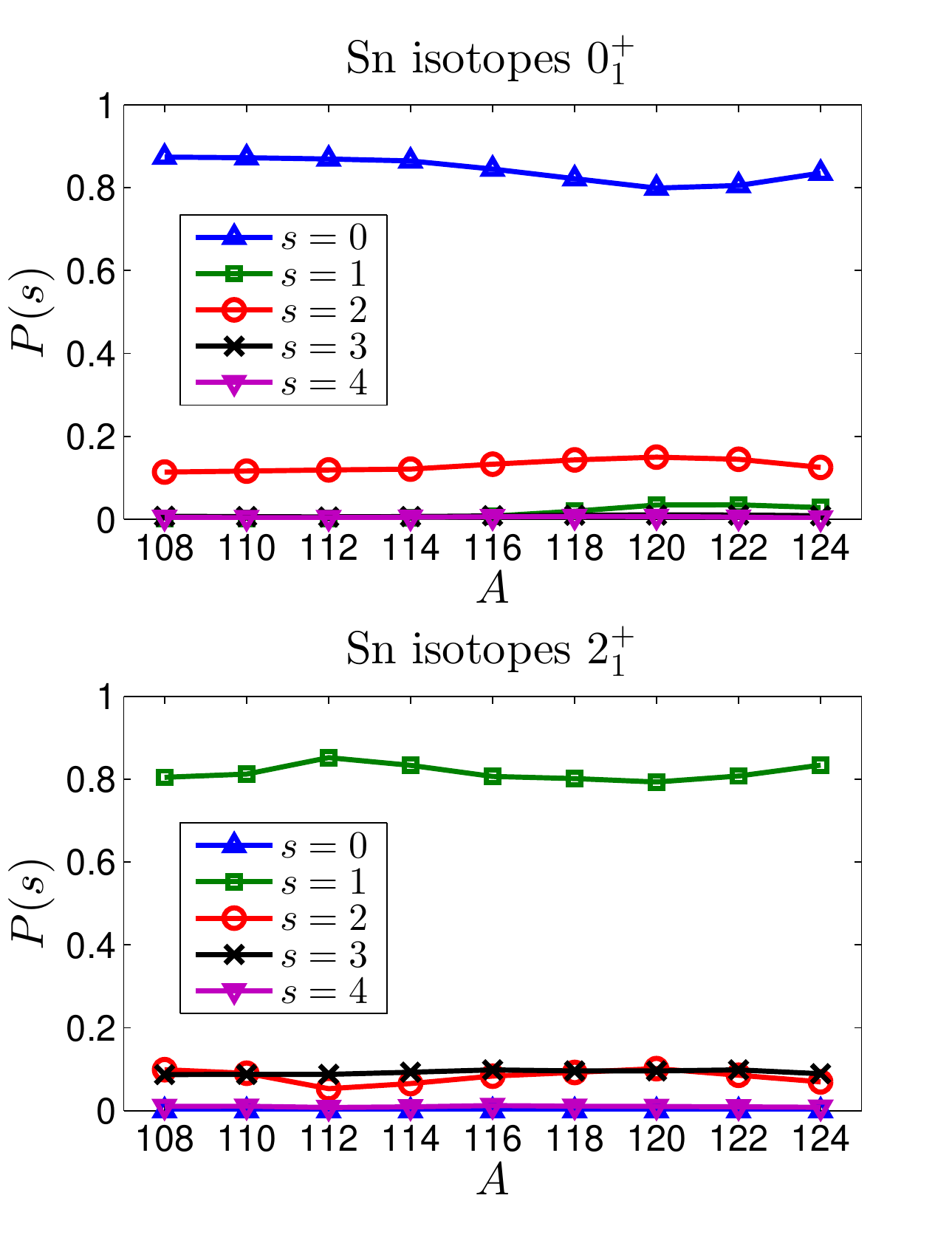}
\caption{\label{Fig_Ps_J0J2} (Color online) Generalized-seniority compositions $P(s)$ of the $0^+_1$ and $2^+_1$ states in Sn isotopes. }
\end{figure}

Figures \ref{Fig_108_sbar}-\ref{Fig_124_sbar} show the generalized-seniority mean $\bar{s}$ (\ref{s_bar}) and fluctuation $\Delta s$ (\ref{Delta_s}) of each eigenstate. The precise composition of the eigenstates could be resolved by the generalized-seniority amplitudes $P(s)$ (\ref{P_s}), as plotted in Figs. \ref{Fig_108_Ps} - \ref{Fig_124_Ps} for $^{108 \sim 124}$Sn. The pattern of successive breakup of the condensed pairs is evident and more pronounced near the mid-shell. In each nucleus, the $P(s=0)$ amplitudes are for a single basis state (\ref{gs}). About $80$ percent of this basis state is saturated into the ground state (see also Fig. \ref{Fig_Ps_J0J2}), and the leftover percent is scattered in many excited $0^+$ states. The $P(s=1)$ amplitudes mainly distribute below the $s=1$ Fermi surface, beyond which the sudden drop is apparent. The $P(s=2)$ amplitudes mostly distribute between the $s=1$ and the $s=2$ Fermi surfaces. Below the $s=1$ Fermi surface the $P(s=2)$ amplitudes are small but not negligible, indicating sizable generalized-seniority mixing into the $s=1$ eigenstates. The $P(s=3)$ amplitudes increase with the excitation energy, but the trend is different for nuclei near and far from the mid-shell. Far from the mid-shell the $P(s=3)$ amplitudes are already large beyond the $s=1$ Fermi surface, and do not increase obviously at the $s=2$ Fermi surface. Whereas near the mid-shell the $P(s=3)$ amplitudes show a clear staircase pattern with sudden increases at the $s=1$ and $s=2$ Fermi surfaces, indicating the persisting superfluid structure. The $P(s=4)$ amplitudes are small and demonstrate the quality of the current calculation truncated to the subspace $|s=4\}$. If generalized seniority was a good truncation scheme, the $P(s)$ amplitudes should decrease with increasing $s$ after the eigen wavefunction achieved convergence. Indeed this is the case here. The $P(s=4)$ amplitudes are small, especially so for low-lying states and for nuclei around the mid-shell. [The $P(s=4)$ amplitudes are not small for $^{108}$Sn and $^{124}$Sn, but the results of these two nuclei are the exact shell-model ones without truncation.] For $^{112 \sim 118}$Sn, the $P(s=4)$ amplitudes are negligible below the $s=1$ Fermi surface and tiny below the $s=2$ Fermi surface, announcing excellent convergence. No exception exists, therefore we should not miss any shell-model eigenstate. The generalized-seniority truncation seems very effective around the mid-shell owing to the best developed superfluid structure; whereas these nuclei have the largest dimension in the standard shell model and are the most time-consuming.

Summarizing Figs. \ref{Fig_108_Ps} - \ref{Fig_124_Ps}, in general the picture of successive breakup of condensed pairs is evident, and more pronounced near the mid-shell owing to the enhanced pairing. At the $s=1$ Fermi surface the $P(s=1)$ amplitudes drop suddenly and the $P(s=2)$ amplitudes increase suddenly, indicating the breakup of the second pair. At the $s=2$ Fermi surface the drop of $P(s=2)$ and the increase of $P(s=3)$ are also evident revealing the breakup of the third pair. The superfluid structure dominates the low-lying spectrum and interprets the wavefunctions. Meanwhile, sizable generalized-seniority mixing exists and the eigenstates do not have pure generalized seniority. Figures \ref{Fig_108_sbar}-\ref{Fig_124_sbar} and Figs. \ref{Fig_108_Ps} - \ref{Fig_124_Ps} display the eigen wavefunctions from different aspects, and they together depict the generalized-seniority pattern.

The near constancy of the first $2^+$ excitation energy has attracted lots of discussions. In fact, it leads Talmi \cite{Talmi_1971, Talmi_book} to propose the generalized seniority as a good quantum number of the Hamiltonian. Talmi assumed the pair structure $v_\alpha$ (\ref{P_dag}) to be invariant along the tin isotopic chain, and studied under what restrictions the $s=0$ state and the $s=1$, $J^\pi = 2^+$ state are eigen states of the Hamiltonian. The derived restrictions lead naturally the constant first $2^+$ excitation energy in the chain. However, the adopted realistic interaction of this work does not fulfill Talmi's picture strictly. In Fig. \ref{Fig_v_in} the pair structures $v_j$ are not constant but vary moderately along the chain. Figure \ref{Fig_Ps_J0J2} shows the generalized-seniority composition $P(s)$ of the ground state $0^+_1$ and the first excited state $2^+_1$. These two states are not pure $s=0$ and $s=1$ states, but have appreciable generalized-seniority mixing. In the ground state $0^+_1$ the dominate $s=0$ component has amplitude $P(s=0) \approx 0.85$, the secondary $s=2$ components have amplitude $P(s=2) \approx 0.13$, and other $s$ components are very small. In the first excited state $2^+_1$ the dominate $s=1$ components have amplitude $P(s=1) \approx 0.82$. Secondarily, the $s=2$ and the $s=3$ components are approximately of equal importance with amplitudes $P(s=2) \approx 0.08$ and $P(s=3) \approx 0.09$. The $P(s=0)$ amplitude vanishes by symmetry and the $P(s=4)$ amplitude is negligible. Although appreciable generalized-seniority mixing exits, we notice that the $P(s)$ compositions of the $0^+_1$ and $2^+_1$ states are almost invariant along the isotopic chain; its origin and possible connection to the constant $2^+$ excitation energy deserve further study.

\begin{figure}
\includegraphics[width = 0.45\textwidth]{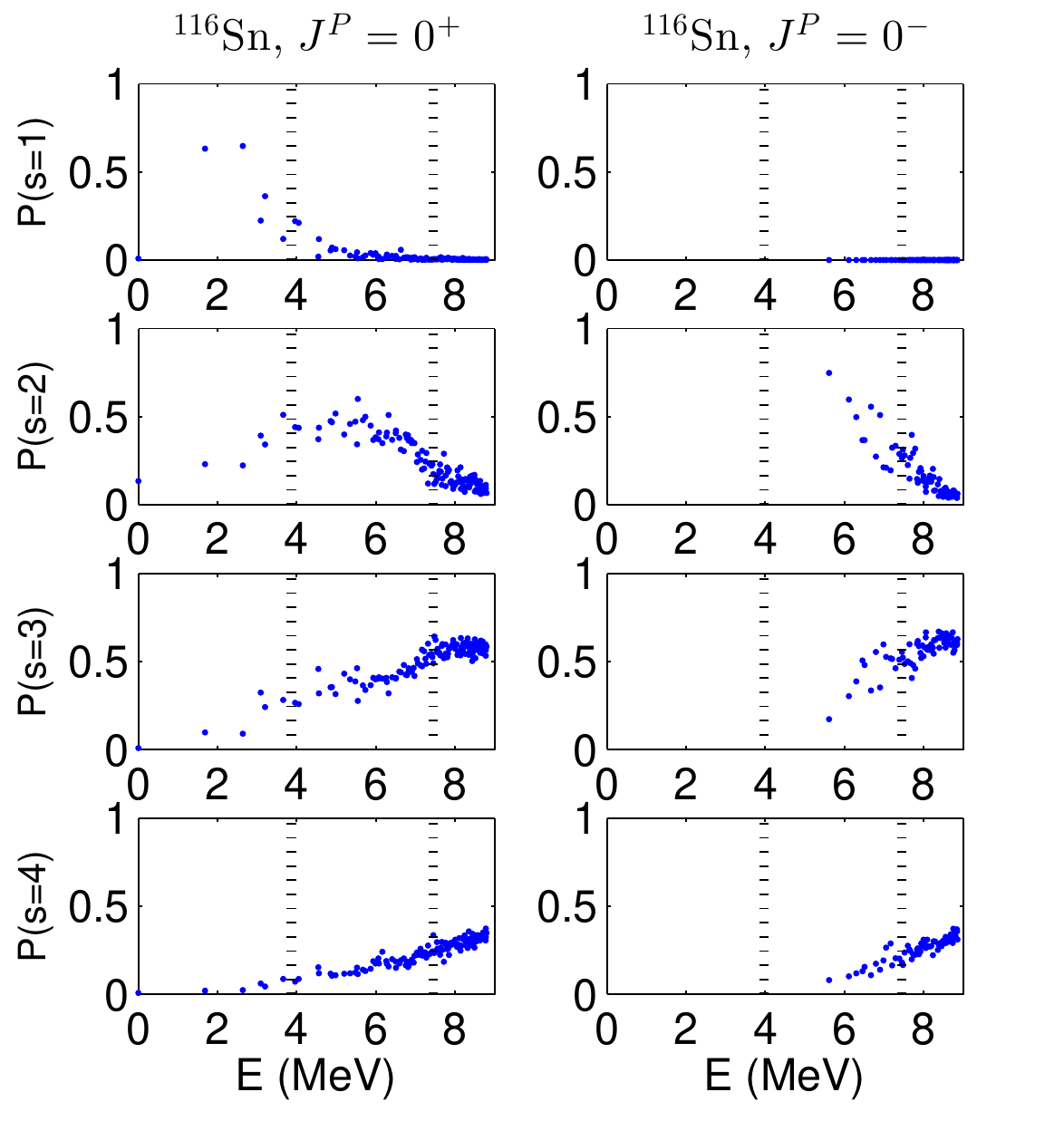}
\caption{\label{Fig_116_J0} (Color online) Generalized-seniority amplitudes $P(s)$ of the $J=0$ states in $^{116}$Sn. In other words we take the data points from those in Fig. (\ref{Fig_116_Ps}) corresponding to $J=0$. The two vertical dotted lines displaying the ``Fermi surfaces'' are at the same position as those in Figs. \ref{Fig_116_sbar} and \ref{Fig_116_Ps}. }
\end{figure}

\begin{figure}
\includegraphics[width = 0.45\textwidth]{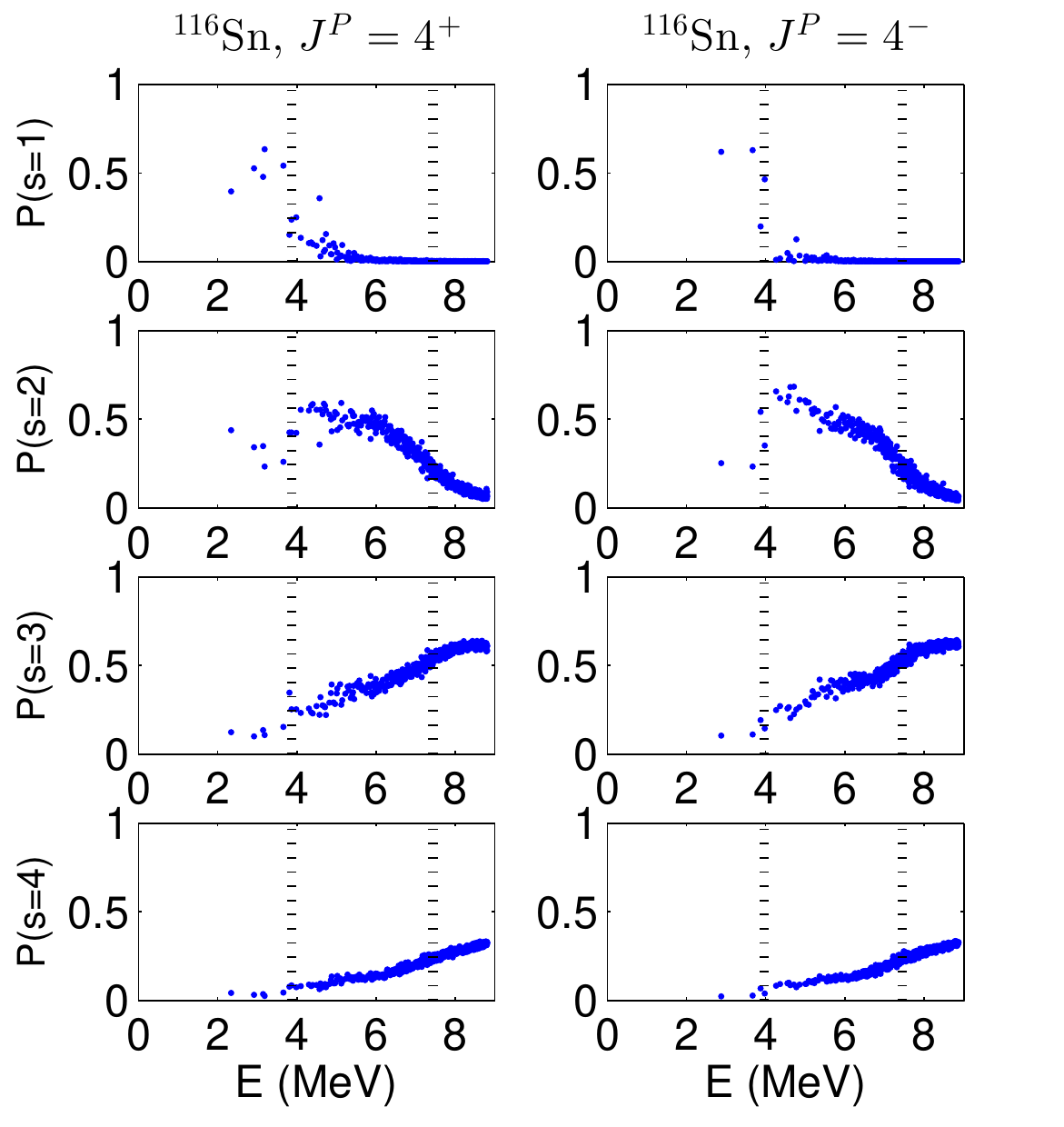}
\caption{\label{Fig_116_J4} (Color online) Generalized-seniority amplitudes of the $J=4$ states in $^{116}$Sn. }
\end{figure}

\begin{figure}
\includegraphics[width = 0.45\textwidth]{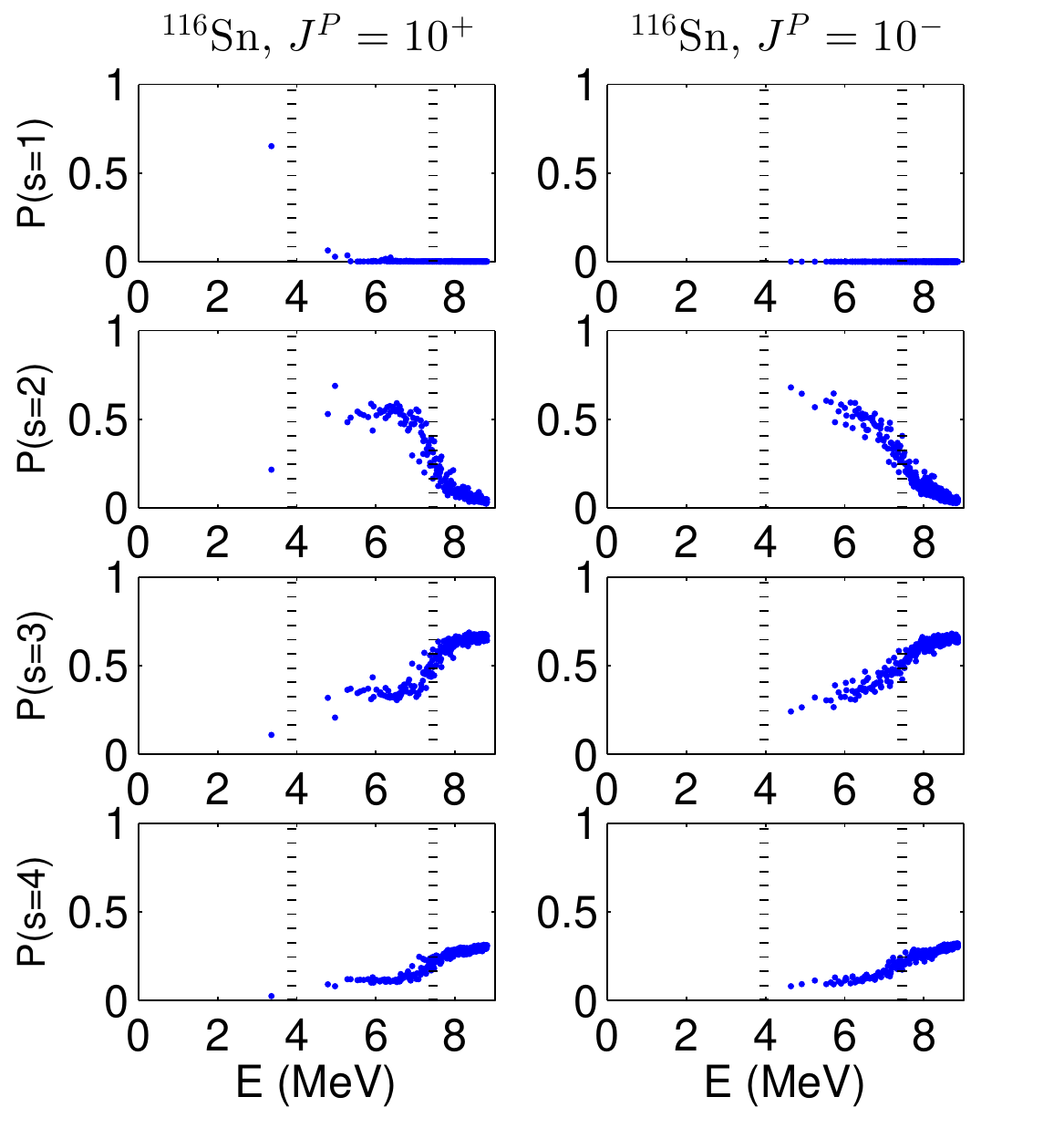}
\caption{\label{Fig_116_J10} (Color online) Generalized-seniority amplitudes of the $J=10$ states in $^{116}$Sn.  }
\end{figure}

Figures \ref{Fig_108_sbar} - \ref{Fig_124_Ps} distinguish the eigenstates only by parity due to the space limitation. More detailed information results from analyzing the eigenstates with the same angular momentum. Here we show six examples of $J^P = 0^+, 0^-, 4^+, 4^-, 10^+, 10^-$ for the mid-shell nucleus $^{116}$Sn in Figs. \ref{Fig_116_J0}, \ref{Fig_116_J4}, and \ref{Fig_116_J10}. We see that the $P(s)$ curve of a given $J^P$ is much narrower (less dispersive at a specific excitation energy) compared with the $P(s)$ curve of a given parity $P$ from Fig. \ref{Fig_116_Ps}. The shapes of the $J=0$, $J=4$, and $J=10$ curves are different; higher $J$ has sharper transitions. The various $J$ curves of different shapes overlap on Fig. \ref{Fig_116_Ps} and result in a more dispersive $P(s)$ curve. Between the $s=1$ and $s=2$ Fermi surfaces the $P(s=4)$ amplitudes are smaller for higher $J$, indicating better convergence. Therefore the generalized-seniority approximation seems better for higher $J$ states, and this deserves further study.

%

%

%

\section{Canonical Ensemble  \label{sec_thermal}}

In this section we compute the mean energy, entropy, and specific heat in the canonical ensemble. Defining $\beta \equiv 1/(k_B T)$, the partition function is
\begin{eqnarray}
Z(\beta) = \sum_i e^{-\beta E_i} = \sum_J (2J+1) e^{-\beta E_J} .  \label{Z_can}
\end{eqnarray}
The probability of occupying the many-body state $i$ is
\begin{eqnarray}
P_i = \frac{ e^{-\beta E_i} }{ Z } .  \nonumber
\end{eqnarray}
The mean energy, mean squared energy, and energy fluctuation are computed as
\begin{eqnarray}
\langle E \rangle = \sum_i E_i P_i ,  \label{E_can} \\
\langle E^2 \rangle = \sum_i (E_i)^2 P_i ,  \label{E2_can} \\
(\Delta E)^2 = \langle E^2 \rangle - \langle E \rangle^2 .  \nonumber
\end{eqnarray}
The entropy and heat capacity are computed as
\begin{eqnarray}
S = k_B \beta \langle E \rangle + k_B \ln Z .  \nonumber \\
C = k_B \beta^2 (\Delta E)^2 .    \nonumber
\end{eqnarray}
For accuracy, we compute quantities by discrete summation rather than taking derivatives of the partition function.

Shell-model type approaches apply two truncations in computing canonical-ensemble quantities: truncating the single-particle basis to the valence space, and the Lanczos diagonalization finds many-body eigenstates up to an energy cutoff $E_c$. In this work the single-particle valence space is the neutron $50 \sim 82$ major shell. This misses the cross-shell excitations at high excitation energies (around $8$ MeV) and brings in errors relative to experiments. The error owing to the Lanczos cutoff $E_c$ is relatively easy to control; we vary $E_c$ and the low-energy part of the results independent of $E_c$ should be reliable.

\begin{figure}
\includegraphics[width = 0.45\textwidth]{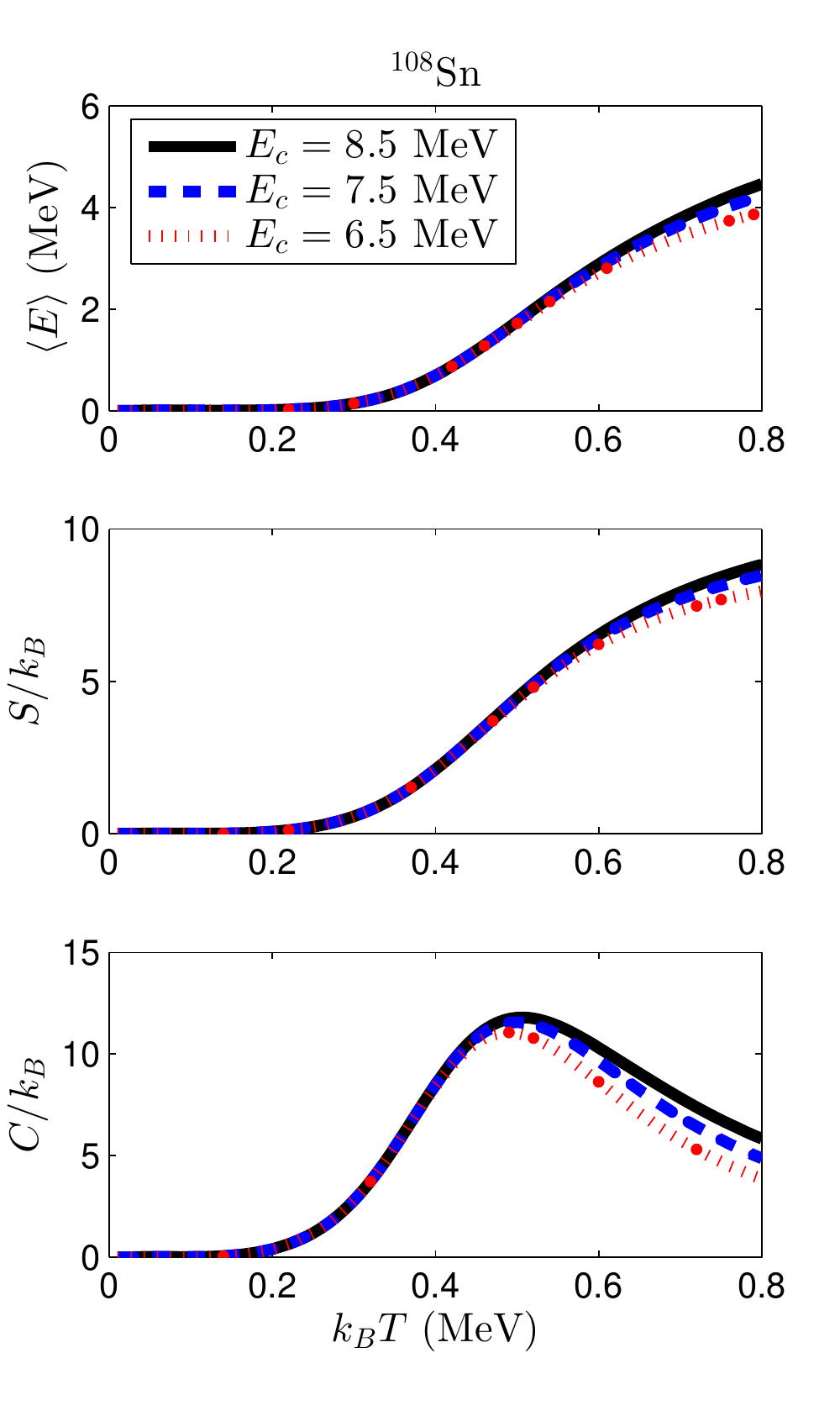}
\caption{\label{Fig_108_ESC} (Color online) The canonical-ensemble mean-energy $\langle E \rangle$, entropy $S$, and specific heat $C$ versus the temperature $T$ in $^{108}$Sn. $k_B$ is the Boltzmann constant. The three curves correspond to the three different energy cutoffs $E_c$. }
\end{figure}

\begin{figure}
\includegraphics[width = 0.45\textwidth]{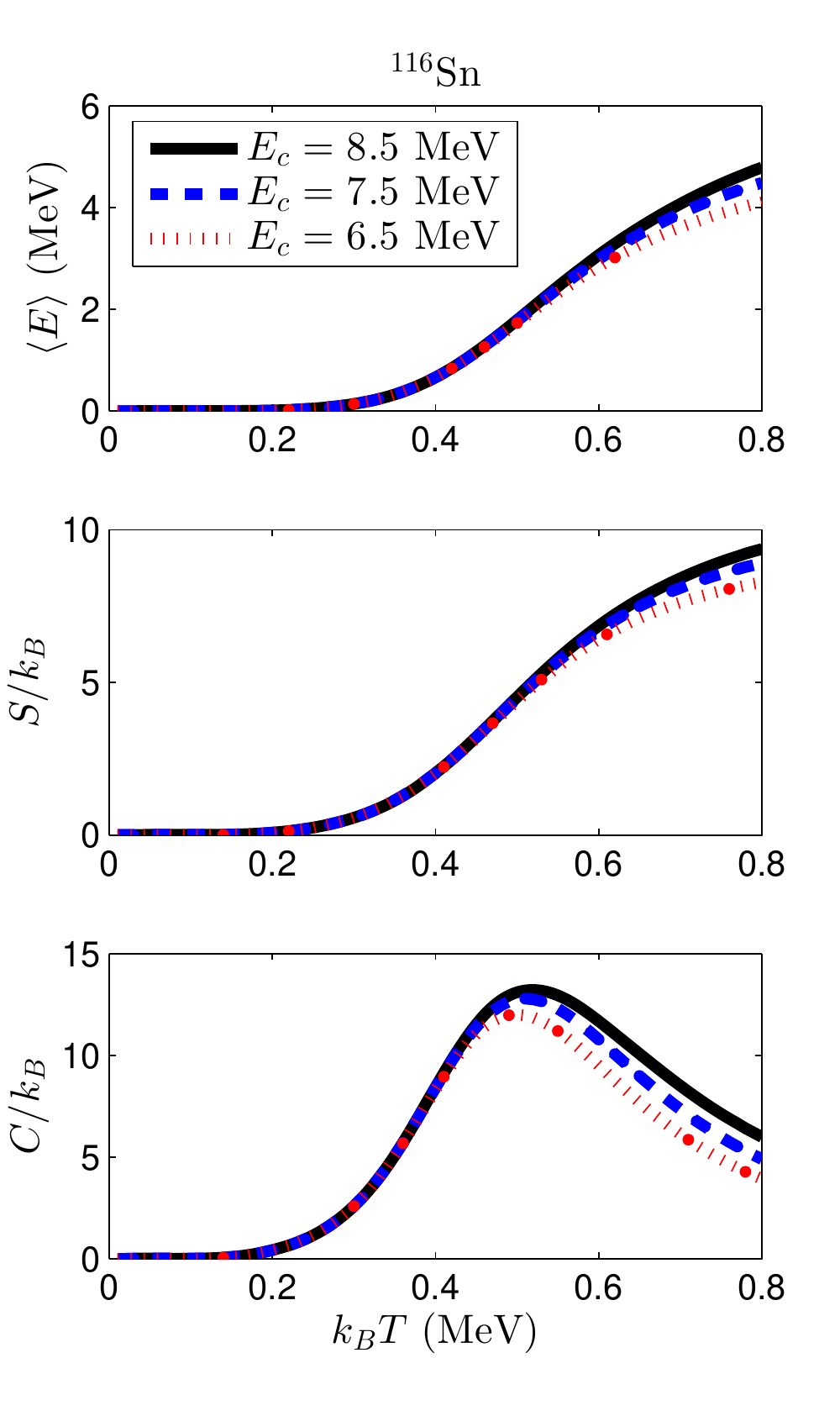}
\caption{\label{Fig_116_ESC} (Color online) The canonical-ensemble mean-energy, entropy, and specific heat versus the temperature in $^{116}$Sn.  }
\end{figure}

\begin{figure}
\includegraphics[width = 0.45\textwidth]{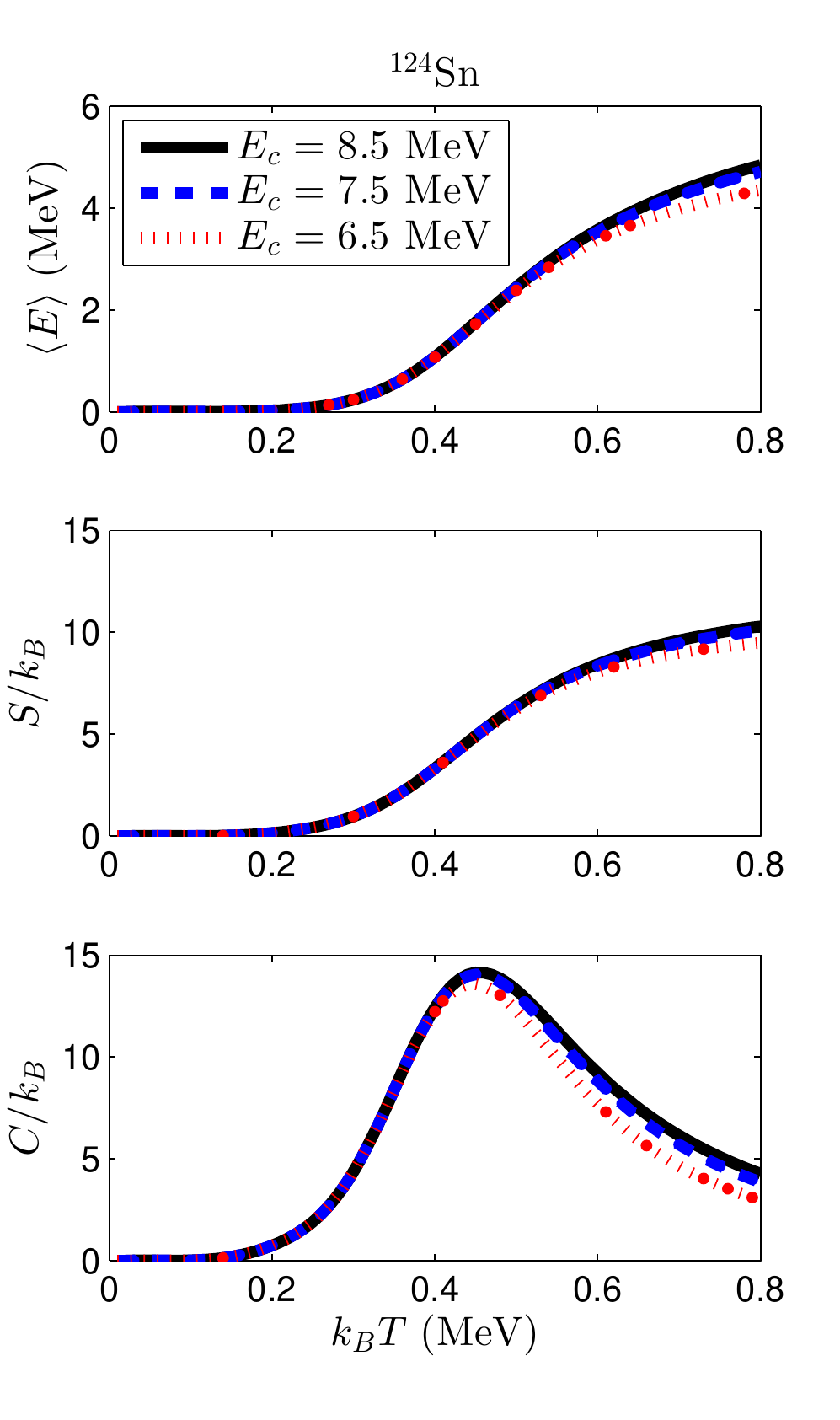}
\caption{\label{Fig_124_ESC} (Color online) The canonical-ensemble mean-energy, entropy, and specific heat versus the temperature in $^{124}$Sn.  }
\end{figure}

We show the results of the canonical-ensemble mean-energy, entropy, and specific heat for three nuclei $^{108}$Sn, $^{116}$Sn, and $^{124}$Sn in Figs. \ref{Fig_108_ESC}, \ref{Fig_116_ESC}, and \ref{Fig_124_ESC}. For each nucleus we take three cutoffs of $E_c = 8.5$, $7.5$, $6.5$ MeV. [In Eqs. (\ref{Z_can}), (\ref{E_can}), and (\ref{E2_can}) the summation includes eigenstates up to $E_c$.] The three corresponding curves below $k_B T = 0.5$ MeV overlap and should be reliable. The pairing phase transition starts around $k_B T_c \approx 0.3$ MeV. The transition temperature $T_c$ in $^{116}$Sn is a little higher than those in $^{108}$Sn and $^{124}$Sn, because of the enhanced pairing around the mid-shell. The steady decrease of the specific heat beyond $k_B T = 0.6$ MeV is unrealistic; the cross-shell excitations not included in the current model space (neutron $50 \sim 82$ major shell) should become important.

\section{Conclusions     \label{sec_conclusions}}

We study even tin isotopes of mass number $A = 108 \sim 124$ with modern realistic interactions in the generalized-seniority truncation of the shell model. Allowing four broken pairs, we compute for each nucleus the lowest $5000$ eigenstates of each parity -- up to around $8$ MeV in excitation energy. The eigen wavefunctions converge well, especially so for low-lying states and for nuclei around the mid-shell. This work promotes the generalized-seniority approximation from ``a viable first
approximation'' \cite{Dean_2003} to an accurate tool of serious realistic calculations for semi-magic nuclei.

The structures of the eigen wavefunctions are investigated in terms of generalized-seniority in detail. For each eigenstate we compute the generalized-seniority ($S=2s$) amplitudes $P(s)$, mean $\bar{s}$, and fluctuation $\Delta s$. The pattern of successive
breakup of the condensed pairs is evident, and
more pronounced near the mid-shell owing to the enhanced collective pairing. Around the mid-shell, the transition is sharp from one to two broken pairs, and is apparent from two to three. The
superfluid structure generated by the pairing force persists at higher energies in the increasingly dense spectrum. The number of eigenstates below the first transition is roughly the same as the dimension of the one-broken-pair subspace, no further dimension truncation is possible. Away from the mid-shell the superfluid structure is more easily destroyed by other correlations at higher energies, but is still useful in interpreting the
eigen wavefunctions.

Meanwhile, sizable generalized-seniority mixing exists even in the mid-shell region and practically no pure generalized-seniority state exists. This suggests more care when using pure generalized-seniority states to describe, for example, the seniority isomers. In particular, the near constancy of the first $2_1^+$ excitation energy may not originate from a generalized-seniority conserving Hamiltonian. However, we observe that the generalized-seniority compositions $P(s)$ of the ground state $0^+_1$ and the first excited state $2^+_1$ are almost invariant along the isotopic chain (see Fig. \ref{Fig_Ps_J0J2}); its origin and possible
connection to the constant $2^+$ excitation energy deserve further study.

We also compute in the canonical ensemble the mean-energy, entropy, and specific heat, based on the $J$-level spectrum up to high excitation energy. The latter is feasible because of the much smaller dimension of the generalized-seniority truncated subspace compared with that of the standard shell model. The thermal pairing phase transition from the superfluid phase to the normal phase is apparent.


%


\section{Acknowledgement}

This work is supported by the National Natural
Science Foundation of China No. 11405109, the Swedish Research Council
(VR) under grant Nos. 621-2012-3805 and 621-2013-4323, the G{\"{o}}ran Gustafsson foundation, and the Hujiang Foundation of China No. B14004. The shell-model calculations were performed on resources provided by the Swedish National Infrastructure for Computing (SNIC) at PDC at KTH, Stockholm.


\newpage
~

\newpage

\clearpage

\end{document}